\documentclass[sigconf]{acmart}
\pdfoutput=1
\usepackage{bm}
\usepackage{url}
\usepackage{bm}
\usepackage{graphicx}
\usepackage{subfigure}
\usepackage{makecell}
\usepackage{algorithm}
\usepackage{algorithmicx}
\usepackage{algpseudocode,caption}
\usepackage{amsmath}
\usepackage{multirow}
\usepackage{enumitem}
\usepackage{caption}
\usepackage{tabularx}
\usepackage{longtable}
\usepackage{supertabular}
\DeclareMathOperator*{\argmax}{argmax}
\usepackage{xspace}

\usepackage{xcolor}  
\usepackage{color}  
\usepackage{colortbl}
\definecolor{shadecolor}{rgb}{0,0,255}
\usepackage{framed} 
\usepackage{url}

\newcommand{\eat}[1]{}
\newcommand{\SeSCE}{\mbox{$\mathsf{UAE}$}\xspace}

\makeatletter
\renewcommand{\@thesubfigure}{\hskip\subfiglabelskip}
\makeatother

\AtBeginDocument{%
  \providecommand\BibTeX{{%
    \normalfont B\kern-0.5em{\scshape i\kern-0.25em b}\kern-0.8em\TeX}}}

\copyrightyear{2021}
\acmYear{2021}
\setcopyright{acmcopyright}\acmConference[SIGMOD '21]{Proceedings of the 2021 International Conference on Management of Data}{June 20--25, 2021}{Virtual Event , China}
\acmBooktitle{Proceedings of the 2021 International Conference on Management of Data (SIGMOD '21), June 20--25, 2021, Virtual Event , China}
\acmPrice{15.00}
\acmDOI{10.1145/3448016.3452830}
\acmISBN{978-1-4503-8343-1/21/06}

\begin{document}
\fancyhead{}

\title{A Unified Deep Model of Learning from both Data and Queries for Cardinality Estimation}

\author{Peizhi Wu$^1$ \qquad Gao Cong$^{1,2}$}

\affiliation{%
  \institution{$\;^1$Singtel Cognitive and Artificial Intelligence Lab for Enterprises@NTU, Singapore}
  \streetaddress{}
  \country{}
  \state{}
  \postcode{}
}
\affiliation{%
  \institution{$\;^2$School of Computer Science and Engineering, Nanyang Technological University, Singapore}
  \streetaddress{}
  \country{}
  \state{}
  \postcode{}
}

\email{{peizhi.wu, gaocong}@ntu.edu.sg}

\begin{abstract}
 Cardinality estimation is a fundamental problem in database systems. To capture the rich joint data distributions of a relational table, most of the existing work either uses data as unsupervised information or uses query workload as supervised information. Very little work has been done to use both types of information, and cannot fully make use of both types of information to learn the joint data distribution.
 In this work, we aim to close the gap between data-driven and query-driven methods by proposing a new unified deep autoregressive model, \SeSCE, that learns the joint data distribution from both the data and query workload. First, to enable using the supervised query information in the deep autoregressive model, we develop differentiable progressive sampling using the Gumbel-Softmax trick. 
 Second, \SeSCE is able to utilize both types of information to learn the joint data distribution in a single model. 
 Comprehensive experimental results demonstrate that \SeSCE achieves single-digit multiplicative error at tail, better accuracies over state-of-the-art methods, and is both space and time efficient.
\end{abstract}

\maketitle

 \vspace{-0.7em}
\section{Introduction} \label{section.1}
Cardinality estimation --- estimating the result size for a SQL predicate --- is  a critical component in query optimization, which 
aims to identify a good execution plan
based on cardinality estimation~\cite{hucost,msdb,postgres}. 
In spite of the importance of cardinality estimation, modern DBMS systems may still produce large estimation errors on complex queries and datasets with strong correlations~\cite{leis2018query}. 
Additionally, cardinality estimation can also be used for approximate query processing~\cite{deepdb}.



%
The fundamental difficulty of cardinality estimation is to construct or learn an accurate and compact representation to estimate the joint data distribution of a relational table (the frequency of each unique tuple normalized by the table's cardinality). 
Most of the existing work on cardinality estimation can be broadly categorized into two classes: data-driven and query-driven cardinality estimation. Data-driven methods aim to summarize the joint data distribution for cardinality estimation. Traditional data-driven methods include data-driven histograms, sampling, and sketching. However, they usually make independence and uniformity assumptions that do not hold in complex real-world datasets. 
%
Learning-based methods have been proposed for data-driven cardinality estimation by formulating cardinality estimation as a machine learning problem. 
Traditional machine learning methods still have their shortcomings. For example, kernel density estimation~\cite{heimel2015self,kiefer2017estimating} is vulnerable to high-dimensional data and probabilistic graphical models~\cite{chow1968approximating, getoor2001selectivity,spiegel2006graph,tzoumas2013efficiently} are inefficient in estimation. 

Recent advances in deep learning have offered promising tools in this regard. 
Recently, the Sum-Product Networks-based model~\cite{deepdb} has been applied  for approximating the joint distribution. However, it will not handle well strong attribute correlations in real-world datasets. 
A promising recent advance in this direction would be applying deep autoregressive models~\cite{nash2019autoregressive,salimans2017pixelcnn++,germain2015made}) for cardinality estimation~\cite{naru, hasan2020deep}, which can capture attribute correlations and have reasonable estimation efficiency. 
%
However, the optimization target (loss function) of deep autoregressive models is to minimize the ordinary \textit{average error} over the overall data, and could neglect tricky (\textit{e.g.,} long tail) data regions, because the model may be largely dominated by those few head data regions, but degraded for many other tail data regions.

Alternatively, query-driven cardinality estimation utilizes query workload (from either query log or generated workload) to perform cardinality estimation without seeing the data, and it is expected to have more focused information on workload queries than data-driven methods~\cite{bruno2001stholes}. Traditional query-driven histograms~\cite{aboulnaga1999self,stillger2001leo,bruno2001stholes,lim2003sash,anagnostopoulos2015learning} also suffer from the same problems of histogram-based methods. 
Recently, deep learning (DL)-based estimators can estimate complex joins without independence assumptions, based on the powerful representation ability of deep neural networks~\cite{schmidhuber2015deep}. However,  query-driven models assume that test queries share similar properties with training queries: they are drawn from the same distribution. This may not be the case due to workload shifts. In other words, test and training workloads may focus on different data regions. Moreover, it would be expensive to generate workload queries that sufficiently cover all data areas to train a model.

It would be a natural idea of utilizing both data and query workload for cardinality estimation. 
In fact, a few proposals (\textit{e.g.,} \textsf{DeepDB}) consider the combination as an interesting avenue for future work. 
Moreover, towards this direction several solutions~\cite{kipf2019learned,dutt2019selectivity,heimel2015self,kiefer2017estimating} have been proposed to utilize both data and workload. 
However, the combination methods of the two types of information in these pioneering studies are simple, and they are not sufficient in capturing both types of information to learn the joint data distribution for cardinality estimation. As to be discussed in related work in more details, these solutions  simply  use one side of data and queries as \textit{auxiliary information} to enhance the model of the other side. Consequently, these pioneering solutions cannot model the data as unsupervised information and query workload as supervised information  in a unified model to learn the joint data distribution for cardinaltiy estimation. 

\noindent{\textbf{Goals}}\quad To solve the aforementioned problems, we conclude four design goals  as follows:
\begin{itemize}[leftmargin=*]
    \item \textbf{G1}. 
    Capturing data correlations without independence or uniformity assumption;
    \item \textbf{G2}. Utilizing both data and query workload for model training;
    \item \textbf{G3}. Incrementally ingesting new data and query workload;
    \item \textbf{G4}. Time and space efficient estimation.
\end{itemize}
\noindent{\textbf{Our Solution}}\quad To achieve the four goals, in this paper we propose a new \textbf{\underline{u}}nified  deep \textbf{\underline{a}}utoregressive \textbf{\underline{e}}stimator \SeSCE to utilize data as unsupervised information and query workload as supervised information for learning the joint data distribution. 
Deep autoregressive models have demonstrated superior performance for their effectiveness  and efficiency in the pioneering work~\cite{naru,hasan2020deep} of training autoregressive models for cardinality estimation. 
However, no existing deep autoregressive model in the literature is able to incorporate the query workload  as supervised information for learning joint data distribution, much less support both data as unsupervised information and query workload as supervised information in the same model.
To enable incorporating query workload as supervised information in the deep autoregressive model to learn the joint data distribution, we propose a novel idea --- we utilize the Gumbel-Softmax trick~\cite{jang2017categorical,maddison2017concrete}  to differentiate the categorically sampled variables so that the deep autoregressive model can learn joint data distribution directly from queries. 
Furthermore, 
we propose to combine the unsupervised and supervised losses produced from data and queries, respectively, with a trade-off hyper-parameter, and thus we are able to train the deep autoregressive model to learn the joint data distribution with a single set of model parameters by minimizing the combined loss. Therefore, \SeSCE can learn from both data and queries simultaneously using the same set of model parameters.
Moreover, since \SeSCE is trained with both data and queries, it is naturally capable of incorporating incremental data and query workload.

\noindent{\textbf{Contributions}}\quad This work makes the following contributions:
\begin{itemize} [leftmargin=*]
    \item We propose a novel approach, \textsf{UAE-Q}, to incorporating query workload as supervised information in the deep autoregressive model.
  This is the first deep autoregressive model that is capable of learning density distribution from query workload.

 \item 
    We propose the \emph{first} unified deep autoregressive model \SeSCE to use both data as unsupervised information and query workload as supervised information for learning joint data distribution. To the best of our knowledge, this is the first deep model that are truly capable of  using query workload as supervised information and data as unsupervised information to learn joint data distribution.

    \item We conduct comprehensive experiments to compare \SeSCE with 9 baseline methods on three real-life datasets. The 9 baseline methods cover data-driven methods, query-driven methods, and hybrid methods, including the recent deep learning based methods. 
    The experimental results show that \SeSCE achieves single-digit multiplicative error at tail, better accuracies over other state-of-the-art estimators, and is both space and time efficient.
    The results demonstrate that our method can well achieve the four aforementioned goals. Interestingly, the experimental results also show that  \textsf{UAE-Q}, which is trained on queries only, outperforms the state-of-the-art query-driven method.

\end{itemize}
\vspace{-1.2em}
\section{Related Work}
Selectivity or cardinality estimation has been an active area of research for decades 
~\cite{survey2012}. 
We present the previous solutions in three categories:  data-driven estimators, query-driven estimators, and hybrid estimators as summarized in Table~\ref{table.models}.



\begin{table*}  
\centering
\begin{center}  
\setlength{\extrarowheight}{.3em}
\renewcommand{\arraystretch}{0.5} 
\resizebox{\textwidth}{33mm}{
\begin{tabular}{|c|c|c|c|c|c|c|c|}  
\hline  
\thead[c]{\textbf{Category}}&\thead[c]{\textbf{Method}} & \textbf{\thead[c]{Without\\ Assumptions}}& \textbf{\thead[c]{Learning\\ from Data}}& \textbf{\thead[c]{Learning\\ from Queries}} & 
\textbf{\thead[c]{Incorporating\\ Incremental Data}} &\textbf{\thead[c]{Incorporating \\ Incremental\\ Query Workload}} & \textbf{\thead[c]{Efficient\\ Estimation}}\\ 
\hline  
\multirow{5}{*}{\thead[c]{\textbf{Data-driven}}}& \thead[c]{Sampling~\cite{lipton1990practical,haas1994relative,riondato2011vc}}&$\checkmark$  &$\checkmark$ &  & $\checkmark$& & \\ 
\cline{2-8} 
& \thead[c]{Histograms~\cite{poosala1996improved, deshpande2001independence,ilyas2004cords,lynch1988selectivity,muralikrishna1988equi,van1993multiple,jagadish2001global,thaper2002dynamic,to2013entropy}} &  &$\checkmark$ &  & & &$\checkmark$  \\  
\cline{2-8} 
 &\thead[c]{KDE~\cite{gunopulos2000approximating,gunopulos2005selectivity}} & $\checkmark$ &$\checkmark$ &  &$\checkmark$ & & $\checkmark$ \\  
\cline{2-8} 
& \thead[c]{PGM~\cite{chow1968approximating,getoor2001selectivity,spiegel2006graph,tzoumas2013efficiently}} &  & $\checkmark$  & & $\checkmark$  & &   \\ 
\cline{2-8} 
& \thead[c]{RSPN model~\cite{deepdb}} &  & $\checkmark$ &  & $\checkmark$  & & $\checkmark$  \\ 
\cline{2-8} 
& \thead[c]{DL models~\cite{naru,hasan2020deep}} & $\checkmark$ &$\checkmark$ &  &$\checkmark$ & & $\checkmark$ \\  
 \hline 
 
 \multirow{3}{*}{\thead[c]{\textbf{Query-driven}}}&
 \thead[c]{Histograms~\cite{aboulnaga1999self,stillger2001leo,bruno2001stholes,lim2003sash,anagnostopoulos2015learning}} & $\checkmark$ & & $\checkmark$&  &$\checkmark$ & $\checkmark$ \\  
  \cline{2-8} 
 & \thead[c]{Mixture models~\cite{park2020quicksel}} &  & & $\checkmark$&  & $\checkmark$& $\checkmark$ \\  
\cline{2-8} 
 & \thead[c]{DL models~\cite{wu2018towards,ortiz2019empirical, hasan2020deep,kipf2019learned, sun2019end}} & $\checkmark$ & &$\checkmark$ &  &$\checkmark$ & $\checkmark$ \\  
\hline 

\multirow{3}{*}{\thead[c]{\textbf{Hybrid}}}&
 \thead[c]{Sampling-enhanced ML models~\cite{kipf2019learned}} & $\checkmark$ & &$\checkmark$  & &$\checkmark$ & \\  
\cline{2-8} 
 & \thead[c]{Histogram-enhanced ML models~\cite{dutt2019selectivity}} & $\checkmark$ & &$\checkmark$  & &$\checkmark$ &$\checkmark$  \\  
\cline{2-8} 
 & \thead[c]{Query-enhanced KDE~\cite{heimel2015self,kiefer2017estimating}} & $\checkmark$  &  $\checkmark$& & $\checkmark$ &  $\checkmark$ &  $\checkmark$ \\  
\cline{2-8} 
 & \thead[c]{\SeSCE (Ours)} & $\checkmark$ &$\checkmark$ &$\checkmark$ &$\checkmark$ & $\checkmark$& $\checkmark$  \\  
\hline 
\end{tabular}}
\caption{A summary of existing cardinality estimation methods.}\label{table.models} 
\end{center} 
\vspace{-4ex}
\end{table*}

\noindent {\textbf{Data-driven Cardinality Estimation}}
Data-driven cardinality estimation methods construct estimation models based on the underlying data.
First, sampling-based methods~\cite{lipton1990practical,haas1994relative,riondato2011vc} estimate cardinalities by scanning a sample of 
data, which has space overhead and can be expensive. Histograms~\cite{poosala1996improved, deshpande2001independence,ilyas2004cords,lynch1988selectivity,muralikrishna1988equi,van1993multiple,jagadish2001global,thaper2002dynamic,to2013entropy} construct histograms to approximate the data distribution. However, most of these methods make partial or conditional independence and uniformity  
assumptions, \textit{i.e.,} the data is uniformly distributed in a bucket.
A host of unsupervised machine learning based methods have been developed for data-driven cardinarlity estimation. 
Probabilistic graphical models (\textsf{PGM})~\cite{chow1968approximating, getoor2001selectivity,spiegel2006graph,tzoumas2013efficiently}  use Bayesian networks to model the joint data distribution, which also relies on conditional independence assumptions.
Kernel density estimation (\textsf{KDE})-based methods~\cite{gunopulos2000approximating,gunopulos2005selectivity} 
do not need the independence assumptions, but their accuracy is not very competitive
due to the difficulty in adjusting the bandwidth parameter. Recently, \textsf{Naru} ~\cite{naru} and \textsf{MADE}~\cite{hasan2020deep} utilize 
unsupervised deep autoregressive models for 
learning the conditional probability distribution and use it
for answering point queries. \textsf{Naru} uses progressive sampling and \textsf{MADE} uses
adaptive importance sampling algorithm for answering range queries and they achieve comparative results. 
Both  \textsf{Naru} and \textsf{MADE}  do not make  any independence assumption.

\noindent {\textbf{Query-driven Cardinality Estimation}}
Supervised query-driven cardinality estimation approaches build models by leveraging the query workload. As opposed to data-driven histograms, query-driven histograms~\cite{aboulnaga1999self,stillger2001leo,bruno2001stholes,lim2003sash,anagnostopoulos2015learning} build histogram buckets from query workload, without seeing the underlying data. 
Recently, \textsf{QuickSel}~\cite{park2020quicksel} uses uniform mixture model to fit the data distribution using every query in the query workload, which avoids the overhead of multi-dimensional histograms.  \textsf{QuickSel}  also relies on uniformity assumptions. %
Deep Learning models have recently been employed for query-driven cardinality estimation.
Ortiz et al.~\cite{ortiz2019empirical} evaluate the performance of using multi-layer perceptron neural networks and recurrent neutral networks on encoded  queries for cardinality estimation. 
\textsf{Sup}~\cite{hasan2020deep} encodes queries  as a set of features and learns weights for these features utilizing
a 
fully connected neural network to estimate the selectivity.
In addition, Wu et al.~\cite{wu2018towards} consider a relevant but different problem, which is to estimate the cardinality for each point of a query plan graph by training a traditional machine learning model. 
Sun et al.~\cite{sun2019end} consider 
estimating both the execution cost of a query plan and caridinality together using a multi-task learning framework. 

\noindent{\textbf{Hybrid Cardinality Estimation}}
%
A few proposals leverage both  query workload and the underlying data to predict the cardinalities. 
Query-enhanced $\mathsf{KDE}$ approaches~\cite{heimel2015self,kiefer2017estimating} leverage query workload to further adjust the bandwidth parameter of \textsf{KDE} to numerically optimize a \textsf{KDE} model for better accuracy. However, \textsf{KDE}-based models do not work well for high-dimensional data~\cite{heimel2015self,kiefer2017estimating}.
%
Recently, selectivity estimation results from data driven models are used  together with encoded queries as input features to 
machine learning models~\cite{kipf2019learned,dutt2019selectivity}. Dutt et al. include the cardinality estimates of histograms as extra features in addition to query features, and use neural network and tree-based ensemble machine learning models for cardinality estimation. 
Kipf et al. ~\cite{kipf2019learned} include estimator results from sampling as extra features in addition to query features and use convolutional neural networks for cardinality estimation. 
However, the two approaches have the following problems: (1)~They cannot be trained on data directly and do not fully capture the benefits of the two types of information;  (2)~Their combination methods significantly increase the model budgets (for storing samples or histograms) and negatively affect the training and estimating efficiencies of the model; (3) They cannot directly ingest incremental data because  they have to be  trained with new queries whose cardinalities are obtained on the updated  data.

\noindent{\textbf{Autoregressive Models}}  Autoregressive models capture the joint data distribution $P(x)$ by decomposing it into a product of conditional distributions. Recent deep autoregressive models include Masked Autoencoder~\cite{made}, Masked Autoregressive Flow~\cite{papamakarios2017masked} and Autoregressive Energy Machines~\cite{nash2019autoregressive}. 


\noindent{\textbf{Remark}} Our  \SeSCE belongs to the hybrid family and is based on deep autoregressive models. To our knowledge, no existing work on  deep autoregressive models in the machine learning literature is able to support using query workload as supervised information to train the model, much less supporting both data as unsupervised information and query workload as supervised information in one model. In \SeSCE, we propose a novel solution to enable using query workload as supervised information, as well as a unified model to utilize both  data as unsupervised information and query workload as supervised information \textit{w.r.t.} deep autoregressive models. 

\vspace{-1.3em}

\section{PROBLEM Statement} \label{section.ps}

Consider a relation $\mathrm{\bm{T}}$ that consists of $n$ columns (or attributes) $\{A_1, A_2,...,A_n\}$. A tuple (or data point) $x \in \mathrm{\bm{T}}$ is an $n$-dimensional vector. The row count of $\mathrm{\bm{T}}$ is defined as $|\mathrm{\bm{T}}|$. The domain region of attribute $A_i$ is given by $R_i$, which represents the set of distinct values in $A_i$.  
~\\
\noindent \textbf{Predicates} A query is a conjunction of predicates, each of which contains an attribute, an operator and a value. A predicate indicates a constraint on an attribute  (e.g., equality constraint  $A_2=6$, or range constraint $A_1> 1$). 

\noindent \textbf{Cardinality} The cardinality of a query $q$, $\mathrm{Card}(q)$, is defined as the number of tuples of $\mathrm{\bm{T}}$ that satisfy the query. Another related term, selectivity, is defined as the fraction of the rows of $\mathrm{\bm{T}}$ that satisfy the query, \textit{i.e.,} ${\rm Sel}(q)= {\rm Card}(q)/|{\rm T}|$. 





\noindent \textbf{Supported Queries} Our proposed estimator supports cardinality estimation for queries with conjunctions of predicates. Each predicate contains a range constraint ($\neq, >, \geq, <, \leq$), equality constraint (=) or $\mathsf{IN}$ clause on a numeric or categorical column. For a numerical column, we make the assumption that the domain region is finite and use the values present in that column as the attribute domain. Moreover, the estimator can also support disjunctions via the inclusion-exclusion principle. Note that our formulation follows a large amount of previous work on cardinality estimation~\cite{naru,hasan2020deep,getoor2001selectivity,park2020quicksel}. For joins, \textsf{UAE} supports multi-way and multi-key equi-joins, as it is  in~\cite{yang2020neurocard}. Moreover, group-by queries could be supported by learning query containment rates \cite{hayek2019improved}.


\noindent \textbf{Problem} 
 Consider (1) the underlying tabular data $\mathrm{\bm{T}}=\{x\}$ and (2) a set of queries $\mathrm{\bm{Q}}=\{ q\}$ with their cardinalities $\mathrm{\bm{C}}=\{ c\}$. Then, this work aims to build a model that 
 leverages the set of queries and their cardinalities $(\mathrm{\bm{Q}}, \mathrm{\bm{C}})$ and the underlying data $\mathrm{\bm{T}}$ to predict the cardinalities for incoming queries.
%
Furthermore, after training the model, it is also desirable that the model can ingest new data and query workload in an incremental fashion, rather than retraining.
Note that such labeled
queries can be collected as feedback from prior query executions (query log). 

\noindent \textbf{Formulation as Distribution Estimation.}
Consider a set of attributes $\{A_1,...,A_n\}$ of a relation $\mathrm{\bm{T}}$ and an indicator function $I(a_1,...,a_n)$ for a query $q$, which produces $1$ if the tuple $(a_1,...,a_n)\in A_1\times...\times A_n$ satisfies the predicate of $q$, and 0 otherwise. The joint data distribution of $\mathrm{\bm{T}}$ is given by $P(a_1,...,a_n)=\frac{\mbox{number of occurrences of } (a_1,...,a_n)}{|\mathrm{\bm{T}}|}$, which is a valid distribution. Next, we can form the selectivity as: ${\rm Sel(q)}=\sum_{x=(a_1,...,a_n)\in A_1\times...\times A_n}{I(x)\cdot P(x)}$. Thus, the key problem of selectivity estimation is obtaining the joint data distribution $P(x)$ under the formulation.

\vspace{-0.8em}
\section{Proposed Model}

We present the proposed \underline{u}nified deep \underline{a}utoregressive \underline{e}stimator, called \textsf{UAE}, that is capable of learning from both data and query workload to support cardinality estimation. 
 We first present an overview of \SeSCE (Section~\ref{subsection:41}). 
 Then we introduce how to use a trained autoregressive model for cardinality estimation (Section~\ref{subsection:42}). We then present our idea on differentiating progressive sampling to enable the deep autoregressive models to be trained with query workload (Section~\ref{subsection:43}). Next, a hybrid training procedure (Section~\ref{subsection:44}) is proposed to use data as unsupervised information and queries as supervised information to jointly train \textsf{UAE}. We present the approaches 
 to incorporating incremental data and query workload in Section~\ref{subsection:45}.
Finally, we discuss several miscellaneous issues (Section~\ref{subsection:4_mi}), and make several remarks (Section~\ref{subsection:46}).
 \vspace{-0.8em}
 
\subsection{Overview} \label{subsection:41}
\begin{figure}[!t]
\centering
\includegraphics[height=3.8cm]{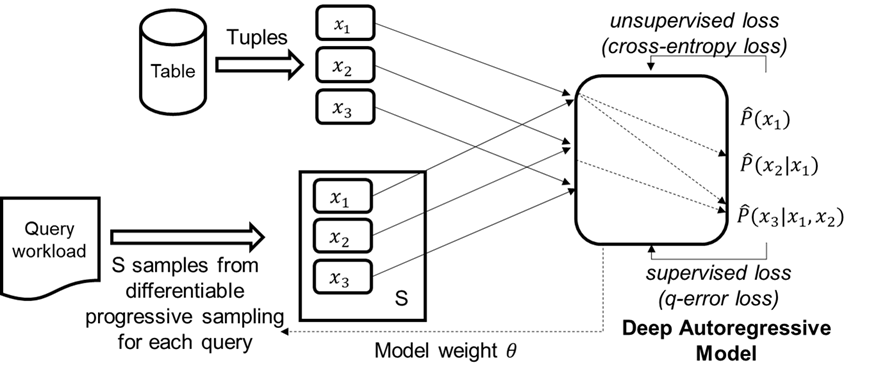}
\vspace{-2.2em}
\caption{Overview of \SeSCE 
}\label{fig.overview}
\vspace{-3ex}
\end{figure}

\noindent {\bf Motivations} 
On the one hand, data-driven methods have been claimed to be more general and robust to workload shifts than query driven methods~\cite{deepdb,naru}. On the other hand,  query workload with true selectivities provides additional information of the workload~\cite{bruno2001stholes}. 
Therefore, it would be a natural idea of combining data-driven and query-driven models. 
As discussed before, the existing proposals leveraging both data and query workload~\cite{kipf2019learned,dutt2019selectivity,heimel2015self,kiefer2017estimating} are insufficient towards this direction.

An idea to overcome the problem of data-driven methods suffering the tail of the distribution due to their averaging optimization target would be using ensemble methods with each component targeting a different part of the distribution. However, 1) it is not easy to define a good partition. 2) It is nontrivial to integrate the results of different ensembles since queries may span multiple ensembles. For example, \cite{deepdb} uses an SPN to combine different ensembles and consequently independence assumptions are made. 3) Using ensembles is orthogonal to \textsf{UAE}. We can integrate \textsf{UAE} with ensemble methods if good ensemble methods could be designed.

\noindent {\bf Challenges} In this work, to achieve the four goals in Introduction, 
 we resort to deep autoregressive models since they ~\cite{naru,hasan2020deep} have demonstrated superior performance for their expressiveness in capturing attribute correlations and efficiency. 
 %
 This is however challenging in two aspects: (1)~Off-the-shelf deep autoregressive models in the machine learning literature are not able to incorporate the query workload information as supervised information for training the model. (2)~ As the naive combination of data-driven and query-driven models is not desirable, we aim to develop a unified autoregressive model with a single set of model parameters to use both data as unsupervised information and query workload as supervised information to learn the joint data distribution. 
 
 
\noindent {\bf Overview of High-level Idea} Both challenges call for designing new deep autoregressive models. 
 First, deep autoregressive models rely on sampling techniques to answer range queries~\cite{naru}. However, it cannot be trained with queries because the sampled categorical variables are not differentiable (to be explained in detail in Section~\ref{subsection:43}). Therefore, to enable the deep autoregressive model to incorporate query workload as supervised information to learn the joint data distribution, we propose a novel idea that we utilize the Gumbel-Softmax trick to differentiate the sampled variables so that the deep autoregressive model can learn the joint data distribution directly from queries. 
 In this way, our proposed model can also incorporate incremental query workload as discussed later.
 
 
Second, to fully leverage data as unsupervised information and queries as supervised information in the hybrid training setting, we combine the unsupervised and supervised losses produced from data and queries, respectively, with a trade-off hyper-parameter. This enables \SeSCE to jointly train the deep autoregressive model to learn the joint data distribution by minimizing the combined loss. Therefore, the deep autoregressive model can learn from both data and queries simultaneously using the same set of model parameters.

 Figure~\ref{fig.overview} shows the workflow of our proposed estimator \SeSCE. We can train \SeSCE with data only, and batches of random tuples are fetched from the table $\mathrm{\bm{T}}$ for learning the joint data distribution. We can also train \SeSCE with query as supervised information only and batches of random (query, cardinality) pairs are read from the query workload log to learn the joint data distribution. \SeSCE is able to lean the joint data distribution with a single autoregressive model from both data and queries. 
  \vspace{-1em}



\subsection{Preliminary: Deep Autoregressive Models for Cardinality Estimation}\label{subsection:42}
\noindent {\bf Autoregressive Decomposition} \quad
Naively, one could store the point distribution of all tuples $(a_1,...,a_n)\in A_1\times...\times A_n$ in a table for exact selectivity estimation. However, the number of entries in the table will grow exponentially in the number of attributes and is not feasible. Many previous methods have attempted to use Bayesian Networks (BN) for approximating the joint distribution $P(x)$ via \textit{factorization}~\cite{getoor2001selectivity,tzoumas2011lightweight}. However, (1) they still make some conditional independence assumptions and (2) the expense of learning and inferring from BN is often prohibitive (we empirically found that the estimation time of a BN could be 110-120s on the DMV dataset).


To achieve a better trade-off between the ability to capture  attribute correlations and space budgets while keeping the tractability and efficiency in model training and inference, we utilize the autoregressive decomposition mechanism which factorizes the joint distribution $P(x)$ in an autoregressive manner without any independence assumption:
\vspace{-1em}
\begin{equation}
    P(a_1,a_2,...,a_n) = \prod_{i=1}^{n}P(a_i|a_1,a_2,...,a_{i-1}).
    \vspace{-0.7em}
\end{equation}
After training on the underlying tuples using neural network architectures, only model weights of the deep autoregressive model need to be stored to compute the conditional distributions $P(A_1), P(A_2|A_1)$ \textit{etc.}. Note that we use left-to-right order in this work, which was demonstrated to be effective in previous work~\cite{naru}. More strategies for choosing a good ordering can be found in~\cite{naru, hasan2020deep}.



\noindent {\bf Encoding Tuples} \quad
Deep autogressive models treat a table as a multi-dimensional discrete distribution. We first scan all attributes to obtain their attribute domain. Next, for each attribute $A_i$, its values are encoded into integers in range $[0, |A_i|-1]$ in a natural order. For instance, consider a string attribute $A_i=\{\textsf{James},\textsf{Tim},\textsf{Paul}\}$, the encoded dictionary would be: $\{\textsf{James}\rightarrow 0,\textsf{Tim}\rightarrow 2,\textsf{Paul}\rightarrow 1 \}$. It is a bijection transformation without any information loss.

After the integer transformation, for each attribute, a specific encoder further encodes these integers into vectors for training the neural networks. The simplest method would be one-hot encoding. Specifically, consider the encoded integers of an attribute $A_i$ with three distinct values: $\{0,1,2\}$, one-hot encoding represents them as $\{(1,0,0), (0,1,0), (0,0,1)\}$. However, this naive method is not efficient in storage because the encoded vector is $|A_i|$-dimensional. We thus use binary encoding, which encodes the same attribute into $\{(00),(01),(10)\}$, a $\log_2{|A_i|}$ dimensional vector. 

\noindent {\bf Model Architectures} \quad We use ResMADE~\cite{nash2019autoregressive}, a multi-layer perceptron with information masking technique which masks out the influence of $x_{\geq i}$ on $\hat{P}(x_i|x_{<i})$. Exploring advanced architectures of deep autoregressive models~\cite{papamakarios2017masked, made} is orthogonal to our work.

\noindent {\bf Model Training} \quad In a nutshell, the input to deep autoregressive estimators is each data tuple and its output is the corresponding predicted density estimation. In the training phase, the weights (or parameters) of deep autoregressive estimators are learned from data tuples by minimizing the cross-entropy between real and estimated data distributions~\cite{germain2015made}.
\begin{equation}\label{eq.LDAEData}
\mathcal{L}^{data} = -\sum_{x\in{\rm T}}{P_\theta(x)\log{\hat{ P}_\theta(x)}}
\end{equation} 
where $\theta$ are the model weights.
Normally, gradient updates for neural networks are performed by stochastic gradient descent (SGD)~\cite{bottou2010large} using \textit{backpropagation}~\cite{rumelhart1987learning,goodfellow2016deep} as a gradient computing technique. Backpropagation is an efficient method for computing gradients in directed graphs of computations, such as neural networks, using \textit{chain rule}~\cite{rudin1964principles}. It is a simple implementation of chain rule of derivatives, which computes all required partial derivatives in linear time in terms of the graph size, as shown in (1) of Figure~\ref{fig.gumbel}. \textit{An important characteristic of backpropagation is that it requires each node of the computation graph involved in the flow to be deterministic and differentiable, which has a well defined gradient.}


\noindent \textbf{Answering Range Queries with Sampling} \quad After being trained, deep autoregressive models can be directly used to answer point queries (\textit{e.g.,} $A_1=2$ AND $A_2=5$ for a relation with two attributes), because a deep autoregressive model is essentially a point distribution estimator.
However, it is not easy to use the point estimator to answer range queries.
 Estimating the selectivity of a range query is equivalent to estimating the sum of selectivities for the set of data points the query contains. Suppose the region of a query $q$ is: $R^q = R^q_1 \times \cdots \times R^q_n$, where $n$ denotes the number of attributes. 
A naive approach for estimating the range query $q$ is exhaustive enumeration:
 \vspace{-0.65em}
\begin{equation}
    \widehat{\rm Sel}(q) = \sum_{x \in R^q}{\hat{P}_\theta(x)},
\end{equation}
 \vspace{-1em}
 
\noindent where $\{x | x \in R^q\}$ represents the list of distinct tuples contained in $R^q$ and $\widehat{\rm Sel}(q)$ is the estimated selectivity of query $q$. However, this method is computationally prohibitive because in the worst case, the number of estimated tuples would grow exponentially in the number of attributes. We thus resort to sampling techniques to efficiently compute the approximate selectivity result as follows. 
 \vspace{-0.5em}
\begin{itemize}[leftmargin=*]
    \item \textbf{Uniform Sampling} method samples $S$ tuples at random and then computes the estimated selectivity as
     \vspace{-0.75em}
    \begin{equation}
    \widehat{\rm Sel}(q) = \frac{|R^q|}{S}{\sum_{s=1}^{S}{\hat{P}_\theta(x^s)}}, 
    \vspace{-0.35em}
    \end{equation}
   
    where $x^s \sim {\rm Uniform(\cdot)}$. However, uniform sampling could produce large variances if the data distribution is skewed.
    \item \textbf{Progressive Sampling} is a Monte Carlo integration approach~\cite{naru}, which sequentially samples each tuple in order of its attributes by \textit{concentrating on the regions of high probability}. Specifically, to sample a tuple $x \in R^q$, we sequentially sample its attributes \{$x_i$\} from distributions \{$\hat P_\theta(\mathbf{X}_i|x_{<i})$\}, respectively, where the categorical distribution $\hat P_{\theta}(\mathbf{X}_i|x_{<i})$ is the distribution of $x_i$ given attributes $x_{<i}$ predicted by the deep autoregressive model. Therefore, the tuple having higher probabilities in $\hat P_{\theta}(\mathbf{X}_i|x_{<i}), (i \in [1,N])$ could be more likely sampled. The selectivity estimate made by a sampled tuple is given by $\hat P_{\theta}(\land \{ \mathbf{X}_i \in R_i^q|x_{<i}\})$. The estimation result from multiple sampled tuples can be easily obtained by averaging the estimate of each single tuple. It is easy to verify that progressive sampling estimates are unbiased. This method is more robust to skewed data distribution than uniform sampling. We thus adopt progressive sampling in our work.
    \end{itemize}
 \vspace{-0.7em}
\subsection{Training Deep Autoregressive Models with Queries}\label{subsection:43}

\begin{figure*}[]
\centering
\vspace{-2ex}
\includegraphics[width=18.5cm,height=7.5cm]{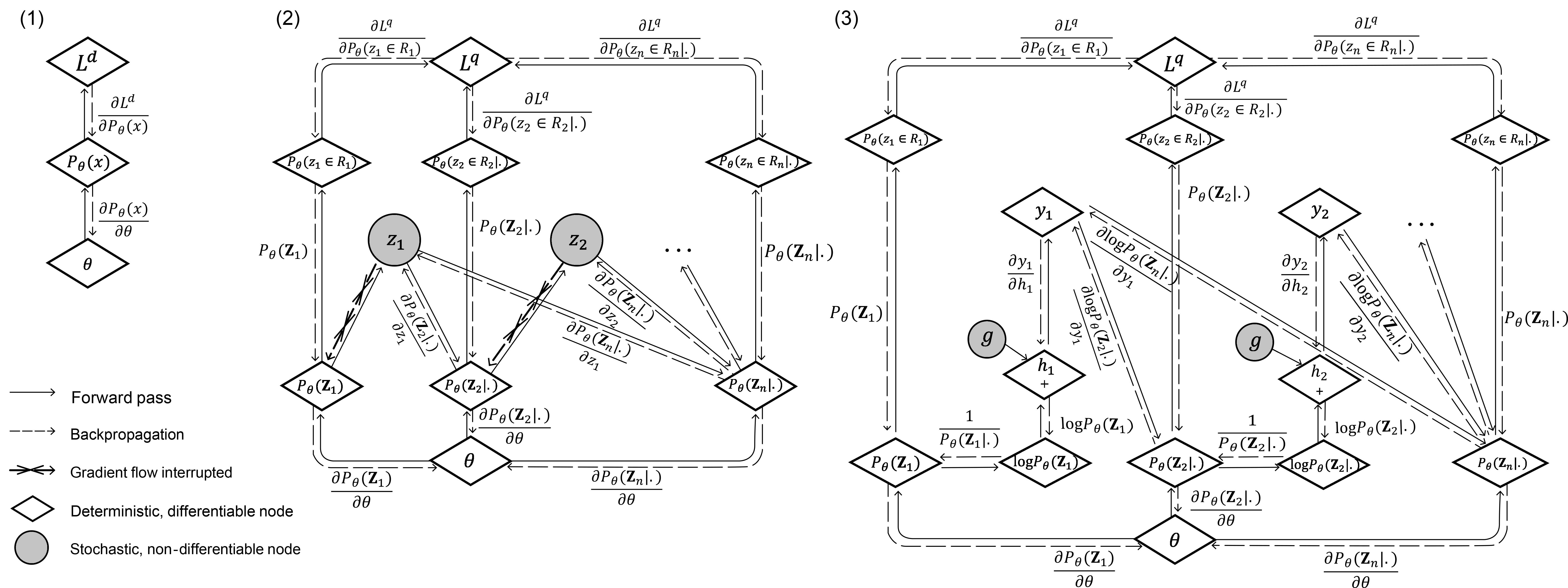}
\vspace{-2.3em}
\caption{\small Gradient estimation in stochastic computation graphs for different models. We denote $\mathcal{L}^{data}$ and $\mathcal{L}^{query}$ by $L^d$ and $L^q$. (1) The autoregressive model trained with data. All nodes are deterministic so gradients can easily flow from $L^d$ to $\theta$. (2) The original autoregressive model fed by queries. The presence of stochastic, non-differentiable nodes $\mathbf{z_1},...,\mathbf{z_{n-1}}$ prevents backpropagation because the categorically sampled variables do not have a well-defined gradient. (3) Our \textsf{UAE} trained with queries. \textsf{UAE} allows gradients to flow from $L^{q}$ to $\theta$ by using a continuous variable $\mathbf{y}$ to approximate $\mathbf{z}$. The stochastic variable $g$ for each attribute is not invloved in the gradient flow from $L^{q}$ to $\theta$.}\label{fig.gumbel}
\end{figure*}


We proceed to present our idea of empowering the autoregressive model with the ability of learning from queries.  
Nevertheless, the existing autoregressive models cannot learn from queries via back-propagation in an end-to-end manner, because in principle gradients cannot flow through the sampled discrete random variables, and hence the process of progressive sampling is not differentiable, which is a prerequisite of backpropagation as explained earlier. Specifically, consider a set of queries $Q=\{q\}$, we define the query loss for autoregressive models as:

 \vspace{-0.9em}
\begin{equation} \label{eq.LDAEQuery}
\mathcal{L}^{query} = \sum_{q \in Q}{\rm Discrepancy}({\rm Sel}(q), \widehat{\rm Sel}(q)),
\end{equation}
 \vspace{-0.2em}
\noindent where $\widehat{\rm Sel}(q)$ is the predicted selectivity of $q$. There are many choices to define the function $\mathsf{Discrepancy(.)}$, e.g., root mean square error (RMSE) and Q-error~\cite{moerkotte2009preventing}:
\begin{equation}
    {\rm Q\text{-}error} = \max {\left(1,  \frac{{\rm Sel}(q)}{\widehat{\rm Sel}(q)}, \frac{\widehat{\rm Sel}(q)}{{\rm Sel}(q)} \right). }\label{eq.qerror}
\end{equation}

Next, let us focus on (2) in Figure~\ref{fig.gumbel}, which illustrates the gradient flow of a deep autoregressive model with progressive sampling using $S=1$ sample, trained with queries. In each forward pass, the autoregressive model utilizes progressive sampling to successively sample one-hot vectors $\mathbf{z_1}, \mathbf{z_2},...,\mathbf{z_{n-1}}$ for each attribute (In practice, the result of sampling from $P_\theta(\mathbf{X}_i|x_{<i})$ in deep autoregressive models is an one-hot vector $z_i$ that represents $x_i$, we thus denote $P_\theta(\mathbf{X}_i|x_{<i})$ as $P_\theta(\mathbf{Z}_i|\cdot)$ thereinafter for clarity) and use them to compute $\widehat{\rm Sel}(q)$. $\mathcal{L}^{query}$ can be obtained after this. However, we observe that during backpropagation, gradients cannot completely flow from $\mathcal{L}^{query}$ to $\theta$. \textit{This is because that gradients cannot flow from $\mathbf{z_1},...,\mathbf{z_{n-1}}$ to $P_\theta(\mathbf{Z_1}),...,P_\theta(\mathbf{Z_{n-1}}|\cdot)$, respectively, since the stochastic variables $\mathbf{z_1},...,\mathbf{z_{n-1}}$ do not have a well-defined gradient \textit{w.r.t.} $P_\theta(\mathbf{Z_1}),...,P_\theta(\mathbf{Z_{n-1}}|\cdot)$.} One can easily generalize the case for $S>1$ as only an averaging operation is needed to combine the estimate of each sample and it does not change the non-differentiable property of progressive sampling. Consequently, the model weights $\theta$ cannot be trained using query workload with the current techniques. 

Our key insight is that if we can find a method making the process of progressive sampling differentiable, the deep autoregressive models can be trained directly from queries by minimizing the discrepancy between the actual selectivities and the estimated selectivities through progressive sampling via back-propagation.

The key challenge of differentiating progressive sampling is \textit{differentiating the non-differentiable sample $\mathbf{z_i}$ from the categorical distribution $P_\theta(\mathbf{Z_i}|\cdot)$}. To this end, we consider two ideas: score function estimators and the Gumbel-Softmax trick, and analyze which is more suitable for our work.

\noindent \textbf{Score Function Estimators.} The \underline{s}core \underline{f}unction estimator (SF), also known as REINFORCE~\cite{williams1992simple}, derives the gradient of query loss for autoregressive models \textit{w.r.t.} the model weights $\theta$ by:
\begin{equation}
\nabla_\theta \mathbb{E}_\mathbf{z}{[\mathcal{L}^{query}]}=\mathbb{E}_\mathbf{z}{[\mathcal{L}^{query}\nabla_\theta{\log P_\theta(\mathbf{Z})} + \nabla_\theta{\mathcal{L}^{query}} ]},
\end{equation}
where $\mathbf{z}$ is sampled from $P_\theta(\mathbf{Z})$, and $\mathcal{L}^{query}$ is a function of $\theta$ and 
$\mathbf{z}$.
With SF, we only need $P_\theta(\mathbf{Z})$ and ${\mathcal{L}^{query}}$ being continuous in $\theta$ (which is valid), without requiring back-propagating through the sampled tuple $\mathbf{z}$.

However, SF often suffers from high variance, even if it is improved with variance reducing techniques~\cite{gu2016muprop}. Also, SF is not scalable if used for categorical distribution because the variance will grow linearly in the number of dimensions in categorical distribution~\cite{rezende2014stochastic}. Consequently, a better method is needed.

\noindent \textbf{The Gumbel-Softmax Trick.} 
\begin{algorithm}[htbp]
    \centering
    \caption{The Gumbel-Softmax Trick (GS-Sampling)}\label{alg.gumbel}
    \begin{algorithmic}[1]
        \Require ~~
        Temperature $\tau$;
        Categorical distribution $\pi$ with $k$ items;
        \Ensure ~~
        A differentiable one-hot vector $\mathbf{e}$ from $\pi$; 
        \State $\text{Sample}\ u \sim {\rm Uniform}(0,1)$ for $j=1,...,k$;
        \State $\text{Compute}\ g_j=-\log{(-\log{(u)})}$ for $j=1,...,k$;
        \State Sample $\mathbf{e}$ according to Eq.\ref{eq.gumbel};
        \State \Return Sampled differentiable one-hot vector $\mathbf{e}$;
    \end{algorithmic}
\end{algorithm}
The Gumbel-Softmax trick~\cite{jang2017categorical,maddison2017concrete}, which was originally used to differentiate discrete latent variables in variational auto-encoders~\cite{kingma2014auto} and is summarized in Algorithm~\ref{alg.gumbel}. We leverage this technique to provide a simple and effective way to draw a sample
$\mathbf{z_i}$ for attribute $A_i$. 

Consider a categorical distribution with $k$-dimensional class probability  $\pi$, to sample a one-hot vector $\mathbf{e}$ from $\pi$, the key idea of the Gumbel-Softmax trick is, 
 \vspace{-0.4em}
\begin{equation}
    \mathbf{e} = {\rm onehot}(\argmax_j \left[g_i+\log{\pi_j} \right]),  \text{for \;} j=1,...,k. 
\end{equation}
 \vspace{-0.2em}
\noindent where $g_1,...,g_k$ are independent and identically distributed samples drawn from a Gumble(0,1) distribution, which can be sampled by: 
 \vspace{-0.3em}
\begin{equation}
g_j=-\log{(-\log{(u)})}, u \sim {\rm Uniform}(0,1). \label{eq.g}
\end{equation}
 \vspace{-0.2em}
Since $\mathsf{argmax}$ is non-differentiable, we can use differentiable $\mathsf{Softmax}$ as a continuous and approximate distribution to sample $\mathbf{e}$:
 \vspace{-0.2em}
\begin{equation}
    \mathbf{e}\approx \mathbf{y}={\rm softmax}({(\log{\pi}+\mathbf{g}})/\tau), \label{eq.gumbel}
\end{equation}
 \vspace{-0.2em}
\noindent where $\tau$ is an adjustable hyper-parameter, referred as temperature. When the temperature $\tau$ approaches 0, sampling from the Gumbel-Softmax distribution becomes one-hot. Hence, the temperature $\tau$ is a trade-off between the gradient variance and the degree of approximation to a one-hot vector. Essentially, $\mathbf{e}$ sampled by the Gumbel-Softmax trick is differentiable and has been proven to have lower variance and be more scalable than SF. Therefore, in this work we use the Gumbel-Softmax trick as the core technique for differentiating progressive sampling. Note that $\pi$ can be any categorical distribution, including $P_\theta(\mathbf{Z_i}|\cdot)$.

Based on the procedure of the Gumbel-Softmax trick, we introduce in detail how to use the Gumbel-Softmax trick to differentiate progressive sampling. As shown in (3) of Figure~\ref{fig.gumbel}, the key idea of differentiable progressive sampling is \textit{using deterministic, continuous variables {$\mathbf{y_1},...,\mathbf{y_{n-1}}$} to approximate stochastic discrete variables { $\mathbf{z_1},...,\mathbf{z_{n-1}}$}} so that gradient can flow from $\mathcal{L}^{query}$ to $\theta$ completely. Specifically, for each attribute $i$, in a forward pass, a stochastic vector $\mathbf{g}$ is first generated from E.q.~\ref{eq.g}. Next, we define $\mathbf{h_i}=(\log P_\theta(\mathbf{Z_i}|\cdot) + \mathbf{g})/\tau$ and sample $\mathbf{y_i}$ from softmax$(\mathbf{h_i})$, according to E.q.~\ref{eq.gumbel}. Note here the categorical distribution $\pi$ in E.q.~\ref{eq.gumbel} is set to $P_\theta(\mathbf{Z_i}|\cdot)$. Then we can use the sampled $\mathbf{y_1},...,\mathbf{y_{n-1}}$ to continue the forward pass. In doing so, we surprisingly find that gradients from $\mathcal{L}^{query}$ to $\theta$ can be computed completely, because the stochastic nodes $\{\mathbf{g}\}$ are dis-encountered from the entire gradient flows.
 \vspace{-0.5em}
\begin{algorithm}[]
    \centering
    \caption{Differentiable Progressive Sampling ($\mathsf{DPS}$)}\label{DPS}
    \begin{algorithmic}[1]
        \Require ~~
        \Statex Temperature $\tau$; Number of samples $S$;
        \Statex Query region $R^q$; Model density estimation $P_\theta(\cdot)$;
        \Ensure ~~
        Estimated selectivity $\widehat{\rm Sel}(q)$; 
        \State $\hat p = 0$; \quad //  {\color{gray}{Initialize the ultimate density estimate}}
        \For{$s=1$ to $S$}
        \State $\hat p^s=1$;\quad //  {\color{gray}{Initialize the $s$-th sample's density estimate}}
        \For{$i=1$ to $n$}
        \State Forward pass the model and obtain $P_\theta(\mathbf{Z_i}|\mathbf{z}_{<i})$;
        \State $\hat p^s= \hat p^s \cdot P_\theta(\mathbf{z}_i \in R_i^q|\mathbf{z}_{<i})$;
        \State Zero-out probabilities outside $R_i^q$ for $P_\theta(\mathbf{Z}_i|\mathbf{z}_{<i})$;
        \State \mbox{Normalize $P_\theta(\mathbf{Z}_i|\mathbf{z}_{<i})$ and obtain $P_\theta(\mathbf{Z}_i|\mathbf{z}_{<i}, \mathbf{z}_i \in R_i^q)$;}
        \State {Sample differentiable $\mathbf{z}_i$ via the Gumbel-Softmax trick: $\mathbf{z}_i$ = GS-Sampling($\tau$, $P_\theta(\mathbf{Z}_i|\mathbf{z}_{<i}, \mathbf{z}_i \in R_i^q)$);}  \quad// {\color{gray}{Alg.~\ref{alg.gumbel}}}
        \EndFor
        \State $\hat p = \hat p + \hat p^s$;
        \EndFor
        \State Average the density estimates of $S$ samples: $\hat p = \hat p / S$;
        \State \Return $ \widehat{\rm Sel}(q) = \hat p$
    \end{algorithmic}
\end{algorithm}
 
We present the flow of differentiable progressive sampling (\textsf{DPS}) in Algorithm~\ref{DPS}. Note that in practice we can perform \textsf{DPS} with $S>1$ samples in batch. Note that in line 7 of Algorithm~\ref{DPS}, we can simply musk out the probabilities of $z_i \notin R_i^q$ by setting the corresponding values in $\log P_\theta(\mathbf{Z_i}|\mathbf{z}_{<i})$ to negative infinity (\textsf{-inf}). This does not change the categorical property of $P_\theta(\mathbf{Z_i}|\mathbf{z}_{<i})$ and does not affect GS-Sampling. 
 \vspace{-1em}

\subsection{Hybrid Training}\label{subsection:44}
\begin{algorithm}[htbp]
    \centering
    \caption{Hybrid Training of \textsf{UAE}}\label{algorithm2}
    \begin{algorithmic}[1]
        \Require ~~
        \Statex Temperature $\tau$; Number of samples $S$; // {\color{gray}{Used in \textsf{DPS}}}
        \Statex $\lambda$; The underlying data $\mathrm{\bm{D}}$; Query workload $(\mathrm{\bm{Q}},\mathrm{\bm{C}})$;
        \Ensure ~~
        Trained model weights $\theta$; 
        \State Randomly initialize model weights $\theta$;
        \While{JointTraining()}
        \State $D_b\gets\text{RandomBatch}(\mathrm{\bm{D}})$;
        \State $(Q_b,C_b)\gets\text{RandomBatch}((\mathrm{\bm{Q}},\mathrm{\bm{C}}))$;
        \State $\text{Obtain}\ \mathcal{L}^{data}\ \text{for}\ D_b$; 
        \State $\text{Obtain}\ \mathcal{L}^{query}\ \text{for}\ (Q_b, C_b)\ \text{through Alg.~\ref{DPS}}\ \text{using}\ \tau,S$;
        \State Perform SGD via backpropagation for $\mathcal{L}$ (Eq.~\ref{eq.loss_all});
        \EndWhile
        \State\Return$\theta$
    \end{algorithmic}\label{sesce-jt}
\end{algorithm} 
Now, \textsf{UAE} is able to learn from either data or queries. 
To achieve our ultimate goal, which is to take both data as unsupervised information and queries as supervised information into the training of \textsf{UAE}, we propose a hybrid training method, which trains the model of \textsf{UAE} by minimizing an overall loss function $\mathcal{L}$ combining $\mathcal{L}^{data}$ (Eq.~\ref{eq.LDAEData}) and $\mathcal{L}^{query}$  (Eq.~\ref{eq.LDAEQuery}) by a hyper-parameter $\lambda$:
 \vspace{-0.5em}
\begin{equation}
    \mathcal{L} = \mathcal{L}^{data}+ \lambda \mathcal{L}^{query}.\label{eq.loss_all}
\end{equation}
 \vspace{-1.2em}
 
The adjustable hyper-parameter $\lambda$ rescales the relative values of $\mathcal{L}^{data}$ and $\mathcal{L}^{query}$. In doing so, \textsf{UAE} learns to capture both the data and query workload information simultaneously to learn the joint data distribution. We summarize the workflow of the hybrid training of \textsf{UAE} in Algorithm~\ref{sesce-jt}.
 \vspace{-1em}

\subsection{Incorporating Incremental Data and Query Workload} \label{subsection:45}

We introduce the superiorities of \textsf{UAE} in efficiently and effectively ingesting incremental data or query workload.

\noindent \textbf{Incremental Data} denotes the tuples newly added to the database after the model is trained. \textsf{UAE} can perform incremental training on the incremental data by minimizing the unsupervised loss $\mathcal{L}^{data}$ produced from the new data. 


\noindent \textbf{Incremental Query Workload} is a set of queries drawn from a different distribution compared to the training workload (\textit{i.e.,} they focus on different data regions). For example, on IMDB dataset, a workload might focus on the data area where \texttt{title.production\_year}>1975 but another workload might focus on \texttt{title.production\_year}<1954. To adapt to the new query workload after being trained, \textsf{UAE} only need to minimize the supervised loss offered from $\mathcal{L}^{query}$ for incrementally ingesting these new queries. In our experiments, we find a smaller value of training epochs (10\textasciitilde20) is enough to prevent \textsf{UAE} from \textit{catastrophic forgetting}.

\vspace{-1em}
\subsection{Miscellaneous Issues} \label{subsection:4_mi}

\noindent \textbf{Supporting Join Queries.} A natural idea of supporting multi-table joins for \textsf{UAE} is to train \textsf{UAE} on join results offered by join samplers~\cite{leis2017cardinality, huang2019joins}. We follow the idea~\cite{yang2020neurocard, deepdb} to handle join queries, which adds \textit{virtual} indicator and fanout columns into the model architecture of \textsf{UAE}. Then we train \textsf{UAE} on tuples sampled by the Exact Weight algorithm~\cite{zhao2018random} and queries with \textit{fanout scaling}. Interested readers may refer to~\cite{yang2020neurocard, deepdb} for details.



\noindent \textbf{Handling Columns with Large NDVs.} A problem of the autoregressive architecture \textsf{UAE}  is when the number of distinct values (NDVs) in a column is very large, storing the model parameters would consume large space. Hence, for columns with large NDVs, we leverage 1) embedding method (which embeds one-hot column vectors by learnable embedding matrices) for tuple encoding and decoding; 2) column factorization which slice a value into groups of bits and then transforms them into base-10 integers~\cite{yang2020neurocard}.

\noindent \textbf{Handling Unqueried Columns (Wildcards).} We use \textit{wildcard skipping}~\cite{liang2020variable, naru} which randomly masks tuples and replaces them with special tokens as the inputs to \textsf{UAE}'s data part during training. 
This could improve the efficiency of training \textsf{UAE} and query inference, because for omitted columns we can skip \textsf{DPS} during training and skip progressive sampling during query inference. 

\vspace{-0.8em}
\subsection{Remarks} \label{subsection:46}

 We call the $\textsf{UAE}$ trained with data and queries as \textsf{UAE-D} and \textsf{UAE-Q}, respectively. We make several remarks as follows:
 
\begin{itemize}[leftmargin=*]
\vspace{-0.4em}
    \item \textsf{UAE} can accurately capture complex attribute correlations without independence or uniformity assumptions because of its deep autoregressive model architecture;
    \item By learning from both data and query workload, \textsf{UAE} is further forced to produce more accurate estimates in the data regions accessed by the workload. Meanwhile, \textsf{UAE} can maintain the knowledge of overall data distribution;
    \item In fact, \textsf{UAE-D} is equivalent to \textsf{Naru}~\cite{naru}. We thus claim that \mbox{\textsf{UAE}} generalizes \textsf{Naru};
    \item We opt for Q-error as the $\mathsf{Discrepancy(.)}$ for \textsf{UAE-Q} because it is consistent with our evaluation metric.
    \item A distinct feature of \textsf{UAE-Q} is that, different from other supervised methods~\cite{wu2018towards,ortiz2019empirical,hasan2020deep,sun2019end} or sampling enhanced ML model~\cite{kipf2019learned} for cardinality estimation which are all discriminative, deep autoregressive model-based \textsf{UAE-Q} is a generative model. To the best of our knowledge, \textsf{UAE-Q} is the \emph{first} supervised deep generative model for cardinality estimation.
    \item When being used to estimate the cadinality of a query, \textsf{UAE} only uses its model weights, without scanning the data. Thus the estimation process is convenient and efficient, especially if accelerated by advanced GPU architectures.
    \item By switching between \textsf{UAE-D} and \textsf{UAE-Q}, \textsf{UAE} can learn from new data or query workload in an incremental manner, without being retrained. To the best of our knowledge, there is no single deep learning-based model for cardinality estimation can achieve the two goals of incremental learning, although it is a consequent advantage of \textsf{UAE}'s construction strategy.
    
\end{itemize}

\vspace{-1em}
\section{EXPERIMENTAL RESULTS} \label{section.exp}
We conduct comprehensive experiments to answer the following research questions.
\vspace{-0.2em}
\begin{itemize}[leftmargin=*]
\item \textbf{RQ1:}\; Compared to state-of-the-art cardinality estimation models, how does \textsf{UAE} perform in accuracy (Section~\ref{subsection:62})?
\item \textbf{RQ2:}\; How different hyper-parameters (\textit{e.g.}, temperature $\tau$, trade-off parameter $\lambda$) affect the results of \textsf{UAE}  (Section~\ref{subsection:63})? 
\item \textbf{RQ3:}\; How well can \textsf{UAE} incrementally incorporate new data and query workload  (Section~\ref{subsection:64})?
\item \textbf{RQ4:}\; How long does it take to train \textsf{UAE}  and how efficient does it produce a cardinality estimate  (Section~\ref{subsection:65})?
\item \textbf{RQ5:}\; How does \textsf{UAE} impact on a query optimizer (Section~\ref{exp.qo})?
\end{itemize}

\subsection{Experimental Settings} \label{subsection:61}

\subsubsection{\textbf{Datasets}}
We use three real-world datasets with different characteristics for single-table experiments as follows:
\begin{itemize}[leftmargin=*]
\item[1.]\textbf{DMV}~\cite{dmv}. This dataset consists of vehicle registration information in New York. We follow the preprocessing strategy in previous work~\cite{naru}, and get 11.6M tuples and 11 columns after preprocessing. 
The 11 columns has widely different data types and domain sizes ranging from 2 to 2101. {\color{black} We also use \textbf{DMV-large} which includes colums with very large NDVs (\textit{e.g.,} 100\% unique \texttt{VIN} column and 31K unique \texttt{CITY} column) 
and has 16 columns. This dataset is used to evaluate the sensitivities of compared methods to very large NDVs. We find that the results provide similar clues as those on \textrm{DMV}, and thus we do not report them here due to the space limit.} 

\item[2.]\textbf{Census}~\cite{uci}. This dataset was extracted from the 1994 Census database, consisting of person income information. It contains 48K tuples and 14 columns. The 14 columns contain a mix of categorical columns and numerical columns with domain sizes ranging from 2 to 123.

\item[3.]\textbf{Kddcup98}~\cite{uci}. This dataset was used in the KDD Cup 98 Challenge. We use 100 columns for experiments and use this dataset to evaluate the sensitivities of various methods to very high-dimensional data . It contains 95K tuples with domain sizes ranging from 2 to 43.

\end{itemize}
\eat{
\begin{table}[htbp]
    \centering
    \begin{tabular}{|c|c|c|}
        \hline
 & \thead[c]{Skewness} & \thead[c]{Correlation} \\
 \hline
 \thead[c]{DMV}& \thead[c]{4.9} &\thead[c]{0.23}  \\
 \hline 
 \thead[c]{Cencus}& \thead[c]{2.1} &  \thead[c]{0.15} \\
 \hline
 \thead[c]{Kddcup98}& \thead[c]{4.7} & \thead[c]{0.32}\\
 \hline
    \end{tabular}
    \caption{{\color{black}Skewness and Correlation of Datasets}}
    \label{table.skew}
    \vspace{-2.3em} 
\end{table}
}

{\color{black} We use two statistics in probability theory to measure the skewness and correlation of datasets: Fisher–Pearson standardized moment coefficient~\cite{doane2011measuring} for skewness and Nonlinear Correlation Information Entropy (NCIE)~\cite{wang2005nonlinear} for correlation. Smaller values of the two measures indicate weaker skewness or correlation. The skewness measures are 4.9, 2.1, 4.7 and the correlation measures are 0.23, 0.15, 0.32 for \textrm{DMV}, \textrm{Census} and \textrm{Kddcup98}, respectively. Therefore, \textrm{DMV} and \textrm{Kddcup98} have relatively stronger skewness and attribute correlation while \textrm{Census} has weaker skewness and attribute correlation.
}
In addition, we use the 
 real-world IMDB dataset for experiments on join queries. IMDB was reported  to have strong attribute correlation~\cite{leis2015good}.

\begin{figure}[htbp]
\vspace{-1.3em}
\flushleft 
\centering
\includegraphics[width=4in]{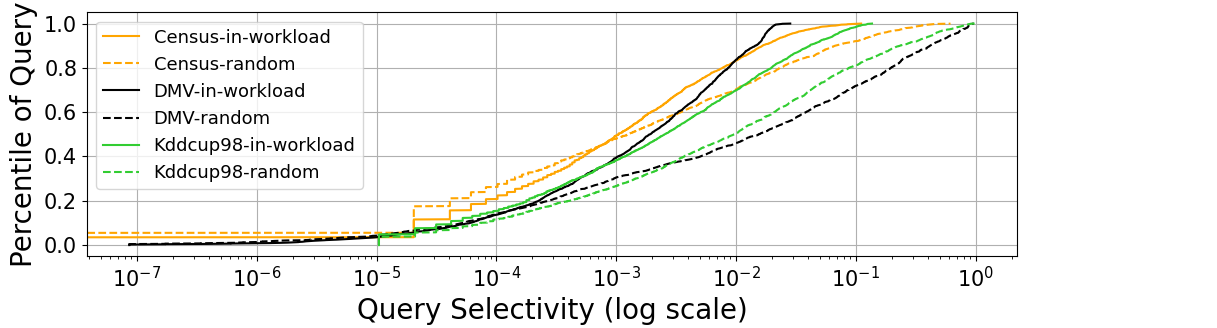}
\centering
\vspace{-1.8em}
\caption{Distribution of Query selectivity.}\label{fig.cards}
\end{figure}

\subsubsection{\textbf{Query Workload}}~\label{exp.workload}\\ 
\noindent\textbf{Training Queries.} 
We follow the previous work~\cite{bruno2001stholes,kipf2019learned} to generate query workload as there is no real query log available for the  datasets we use. Specifically, we first choose an attribute with a relatively large domain size as the bounded attribute. The bounded attribute is specified by a distribution for the centers and a target measure~\cite{bruno2001stholes}, which is based on real usage scenarios. The distribution center is chosen uniformly within a specific range and the target measure is a target volume of 1\% of the distinct values. We have also varied the selection method for the centers (\textit{e.g.,} following data distribution) and the target measure (\textit{e.g.}, target selectivity) and the experimental results turned out to be qualitatively similar. We thus do not report these results due to the page limit. Next, for other attributes (\textit{i.e.}, random attributes), we follow the method~\cite{kipf2019learned, naru} to generate queries. We draw the number of filters ($n_f \geq 5$) at random. Then we uniformly sample $n_f$ columns and the corresponding filter operators. Finally, the filter literals are set from the values of a randomly sampled tuple. We generate 20K training queries for each dataset.
{\color{black} For join experiments, 
we use one template (a join table subset) out of 18 templates in \textsf{JOB-light}, a 70-query benchmark used by a number of previous work, to generate 10K training queries. This template includes 3 tables, \texttt{title}, \texttt{movie\_companies}, \texttt{movie\_info}. We set \texttt{title.production\_year} as the bounded attribute and then choose 2-5 filters on other content columns as discussed above. This generation procedure  follows~\cite{yang2020neurocard} which produces more diversified queries than \texttt{job-light}, using the same join template. We term this benchmark as \textsf{JOB-light-ranges-focused}}, 

\noindent \textbf{Test Queries.} 
Apart from the performance on the training workload (\textit{i.e.}, in-workload queries), we also evaluate whether the estimators are robust to out-of-workload queries. Therefore, we generate two kinds of test queries to thoroughly evaluate the performance of estimators: (1) In-workload Queries: 2K test queries are generated in the same procedure of training query workload. {\color{black} For joins experiments we generate 1K test queries from \textsf{JOB-light-ranges-focused};} (2) Random Queries: We also generate 2K test queries without bounded attributes, \textit{i.e.}, all attribute filters are generated randomly, {\color{black} to evaluate the robustness of different models to workload shifts. For join experiments we use \textsf{JOB-light} as it contains no focused information.

\noindent \textbf{Workload Characteristics.} Figure~\ref{fig.cards} plots the selectivity distributions of 2K in-workload and random queries on all datasets. We observe: 1) the selectivities of generated workloads are widely spaced. 2) Random queries have much wider selectivity spectrums than in-workload queries because in-workload queries have an additional bounded column.
Note that though training and test in-workload queries share the same generation procedure,
we manually ensure that each training query is different from each test query.}
\vspace{-0.5em}
\subsubsection{\textbf{Performance Metric}}  
Following the previous work~\cite{kipf2019learned, naru, hasan2020deep}, we evaluate all models by the popular metric for cardinality estimation, q-error, which is defined in E.q.~\ref{eq.qerror}.
\vspace{-0.5em}
\subsubsection{\textbf{Baseline Methods}}
We compare \SeSCE~\footnote{The source code is available at \url{https://github.com/pagegitss/UAE}.} to 9 cardinality estimation methods, including state-of-the-art and the newest methods.

\noindent\textbf{\underline{Query-driven Models}}:
\begin{itemize}[leftmargin=*]
\item[1.]\textsf{MSCN-base}~\cite{kipf2019learned}. This query-driven deep learning (DL)-based method uses a multi-set convolutional neural network for answering correlated joins. For each predicate, it featurizes the attribute and operator using one-hot vectors and normalizes the value. It then concatenates the average results over the predicate set as the query encoding. We use two layers (256 hidden units) of multilayer perceptrons, the default setting of~\cite{kipf2019learned}, on the query encoding. We apply the code from~\cite{mscncode}. {\color{black}{Note that the original \textsf{MSCN} was proposed to handle join queries. We adapt \textsf{MSCN} to single-table queries by dropping the join module}.} We also evaluate another query-driven DL method \textsf{Sup}~\cite{hasan2020deep} and find that it shares the similar performance with \textsf{MSCN}. 
\item[2.]\textsf{LR}~\cite{kutner2005applied}. This method first represents a query as the concatenation of the domain range of each predicate (following~\cite{dutt2019selectivity}), and trains a linear regression model on the query representation. We use this method as a non-DL query-driven counterpart to demonstrate the effectiveness of DL-based query-driven methods (\textsf{MSCN-base}).
\end{itemize}

\noindent\textbf{\underline{Data-driven Models}}:
\begin{itemize}[leftmargin=*]
\item[3.]\textsf{Sampling}. This method keeps a portion {\color{black}($p$)} of tuples uniformly from the dataset and scans them to estimate query cardinalities. 
\item[4.]\textsf{BayesNet}~\cite{chow1968approximating}. 
We follow the same setting~\cite{naru} for this method for a fair comparison.  
\item[5.]\textsf{KDE}~\cite{gunopulos2005selectivity}. This method leverages the kernel density estimator for estimating the data distribution. Gaussian kernels are adopted in the experiments and the bandwidth is calculated from data by Scott's rule~\cite{scott2015multivariate}.
\item[6.]\textsf{DeepDB}~\cite{deepdb}. This method models joint data distribution by learning relational sum-product networks, which is based on the structure of Sum Product Networks (\textsf{SPN})~\cite{poon2011sum}. The number of samples per \textsf{SPN} for learning the ensemble is set to 1M. We use its open-sourced code~\cite{deepdbcode}. \textsf{DeepDB} is a deep model but non-neural, which is a proxy to compare neural deep models (\textsf{Naru}, \textsf{UAE}) against the effectiveness of non-neural deep models.
\item[7.]\textsf{Naru}~\cite{naru}. 
\textsf{Naru} is equivalent to \textsf{UAE-D}. We extend the open-sourced code from~\cite{narucode} because the original code does not support two-sided queries. We also compare with \textsf{MADE}~\cite{hasan2020deep}, which also uses deep autoregressive models and its performance is close to \textsf{Naru}~\cite{naru}.
{\color{black}{For join queries, we compare \textsf{UAE} with \textsf{NeuroCard}~\cite{yang2020neurocard}, a concurrent work that extends \textsf{Naru} for join}.}
\end{itemize}
\noindent\textbf{\underline{Hybrid Models}}:
\begin{itemize}[leftmargin=*]
\item[8.]\textsf{MSCN+sampling}~\cite{kipf2019learned}. This method uses estimates on materialized sampled tuples as additional inputs to \textsf{MSCN-base}. We use this method to demonstrate the advantages of leveraging both data and workload information.
\item[9.]\textsf{Feedback-KDE}~\cite{heimel2015self}. This method further utilizes query feedback to adjust the bandwidth in \textsf{KDE}~\cite{gunopulos2005selectivity}. We apply the code from the authors~\cite{feedbackkde} and modify it to run it with more than 10 columns. \textit{SquaredQ} loss function and \textit{Batch} variant are adopted for bandwidth optimization.
\end{itemize}
 {\color{black} We also compared with \textsf{STHoles}~\cite{bruno2001stholes}, \textsf{Postgres}~\cite{postgres} and \textsf{MHIST}~\cite{poosala1996improved}. The performances are 
 worse than the 9 methods, and thus we do not report them here.}
The numbers of sample tuples in two \textsf{KDE}-based methods (\textsf{KDE} and \textsf{Feedback-KDE}) and two sampling-based methods (\textsf{Sampling} and \textsf{MSCN+sampling}) are set to match the memory budget of our model for a fair comparison. {\color{black} For \textsf{Sampling} method, the sample ratios are 0.2\%, 9\%, 4.6\% for \textrm{DMV}, \textrm{Census} and \textrm{Kddcup98}, respectively. We also test two sampling budgets, one smaller ($0.75p$) and one lager ($1.25p$) than $p$. The results do not change the conclusions in this paper, and thus we do not report them here.} Additionally, for a fair comparison, two autoregressive based models (\SeSCE and \textsf{Naru}) share the same model architecture, which is 2 hidden layers 
($2\times 128$ units). We also turn on column factorization on IMDB due to its high-cardinality columns. Afterward, the space consumption of the autoregressive model is 2.0MB, 0.5MB, 3.45MB and 4.1MB on DMV, Census, Kddcup98 and IMDB, respectively. 
In addition, the two deep autoregressive based methods, \textsf{Naru} and \textsf{UAE}, rely on progressive sampling for answering range queries. For fair comparison, the number of estimate samples is set to 200 on DMV in-workload queries, Census and 1K on DMV random queries, Kddcup98, IMDB for both of them, because we empirically found by this setting, they can strike a better balance between the estimation accuracy and overheads, \textit{e.g.,} further increasing the estimate number will not result in a significantly improvement in accuracy but will increase the estimation overhead.
%
In \SeSCE, the temperature $\tau$ and the number of samples $S$ in \textsf{DPS} are set to 1 and 200, respectively, on all the datasets. 
The trade-off parameter $\lambda$ is set to $1e^{-4}$ on three single-table datasets and 10 on IMDB dataset.


All the experiments were run on a machine with a Tesla V100 GPU and a 20 cores E5-2698 v4 @ 2.20GHz CPU.


\vspace{-1.2em}
\subsection{Performance Comparison} \label{subsection:62}

\begin{table*}[ht]
  \centering
 \begin{minipage}[t]{0.49\textwidth}
  \centering 
     \makeatletter\def\@captype{table}\makeatother
\scalebox{0.75}{\begin{tabular}{|l|l|l|l|l|l|l|l|l|l|}
  \hline
  \thead[c]{Model}&  \thead[c]{{\color{black}Size}} &  \multicolumn{4}{c|}{\thead[c]{In-workload Queries}} &  \multicolumn{4}{c|}{\thead[c]{Random Queries}}\\
  \cline{3-10} 
 & &  \thead[c]{Mean} &  \thead[c]{Median} &  \thead[c]{95th} &  \thead[c]{MAX} &  \thead[c]{Mean} &  \thead[c]{Median} &  \thead[c]{95th} &  \thead[c]{MAX} \\
  \hline
\thead[c]{LR} &  {\color{black}\thead[c]{{17KB}}}&\thead[c]{85.51} & \thead[c]{3.683}&\thead[c]{21.31} &\thead[c]{$1\cdot10^5$} & \thead[c]{$1\cdot10^6$} & \thead[c]{$2\cdot10^5$} & \thead[c]{$8\cdot10^6$} &\thead[c]{$1\cdot10^7$} \\
 \hline
 \thead[c]{MSCN-base}& {\color{black}\thead[c]{{0.5MB}}} &\thead[c]{5.474} & \thead[c]{2.219}&\thead[c]{9.641} & \thead[c]{2580}& \thead[c]{2537} & \thead[c]{235.3} & \thead[c]{$1\cdot10^4$} & \thead[c]{$7\cdot10^4$} \\
 \hline  
  \thead[c]{\textsf{UAE-Q}}&{\color{black}\thead[c]{{2MB}}}
  &\thead[c]{2.78} & \thead[c]{3.14}&\thead[c]{3.72} & \thead[c]{108.0}& \thead[c]{317} & \thead[c]{1.68} & \thead[c]{68.35} & \thead[c]{$1\cdot10^5$} \\
 \hline \hline
 \thead[c]{Sampling} & {\color{black}\thead[c]{{2MB}}}
 & \thead[c]{24.66}& \thead[c]{1.104}&\thead[c]{65.25} &\thead[c]{2179} &\thead[c]{21.96}  & \thead[c]{1.036} &\thead[c]{56.1}  &\thead[c]{2086} \\
 \hline
 \thead[c]{BayesNet} & {\color{black}\thead[c]{{1.9MB}}} & \thead[c]{1.233}&\thead[c]{1.044} & \thead[c]{1.325}& \thead[c]{174.0}& \thead[c]{3.106} & \thead[c]{1.041} & \thead[c]{9.364} & \thead[c]{480.0}\\
 \hline
 \thead[c]{KDE} &{\color{black}\thead[c]{{2.1MB}}}& \thead[c]{22.48}&\thead[c]{1.299} & \thead[c]{4.302}& \thead[c]{$1\cdot10^4$}& \thead[c]{70.16} &\thead[c]{1.190}  & \thead[c]{46.06} &\thead[c]{$2\cdot10^4$} \\
 \hline
 \thead[c]{DeepDB} &{\color{black}\thead[c]{{1.3MB}}}&\thead[c]{1.454} & \thead[c]{1.120}&\thead[c]{1.519} & \thead[c]{293.5}& \thead[c]{836.3} & \thead[c]{1.058} & \thead[c]{5347} &\thead[c]{$3\cdot10^4$} \\
 \hline
 \thead[c]{Naru} &{\color{black}\thead[c]{{2MB}}}
 & \thead[c]{1.113}&\thead[c]{1.029} &\thead[c]{1.160} & \thead[c]{108.0}& \thead[c]{\textbf{1.062}} & \thead[c]{\textbf{1.021}} & \thead[c]{1.213} & \thead[c]{7.0}\\
 \hline \hline
 \thead[c]{MSCN+sampling} & {\color{black}\thead[c]{{2MB}}}
 & \thead[c]{1.943}&\thead[c]{1.068} & \thead[c]{4.942}& \thead[c]{122.6}& \thead[c]{25.31} & \thead[c]{6.037} & \thead[c]{109.3} & \thead[c]{760.5}\\
 \hline
 \thead[c]{Feedback-KDE} &{\color{black}\thead[c]{{2.1MB}}}& \thead[c]{27.63}& \thead[c]{1.288}&\thead[c]{4.20} &\thead[c]{$1\cdot10^4$} & \thead[c]{110.3} & \thead[c]{1.184} & \thead[c]{54.7} &\thead[c]{2041} \\
 \hline \hline
 \thead[c]{\SeSCE (Ours)}& {\color{black}\thead[c]{{2MB}}} &\thead[c]{\textbf{1.058}}&\thead[c]{\textbf{1.027}} &\thead[c]{\textbf{1.143}} & \thead[c]{\textbf{5.0}}& \thead[c]{\textbf{1.062}} & \thead[c]{1.024} & \thead[c]{\textbf{1.20}} &\thead[c]{\textbf{6.0}} \\
 \hline 
\end{tabular}} \caption{Estimation Errors on DMV}\label{table.2}
\end{minipage}
  \begin{minipage}[t]{0.49\textwidth}
  \centering
        \makeatletter\def\@captype{table}\makeatother
\scalebox{0.75}{\begin{tabular}{|l|l|l|l|l|l|l|l|l|l|}
  \hline
  \thead[c]{Model} & \thead[c]{{\color{black}Size}}&  \multicolumn{4}{c|}{\thead[c]{In-workload Queries}} &  \multicolumn{4}{c|}{\thead[c]{Random Queries}}\\
  \cline{3-10} 
   & & \thead[c]{Mean} &  \thead[c]{Median} &  \thead[c]{95th} &  \thead[c]{MAX} &  \thead[c]{Mean} &  \thead[c]{Median} &  \thead[c]{95th} &  \thead[c]{MAX} \\
  \hline
\thead[c]{LR}& {\color{black}\thead[c]{{14KB}}} & \thead[c]{6.876} & \thead[c]{2.963}& \thead[c]{20.154}&\thead[c]{467.0} & \thead[c]{3949} & \thead[c]{200.2} &  \thead[c]{$3\cdot10^4$}& \thead[c]{$4\cdot10^4$}\\
 \hline
 \thead[c]{MSCN-base}& {\color{black}\thead[c]{{0.3MB}}}& \thead[c]{5.90}&\thead[c]{2.563} &\thead[c]{17.72} & \thead[c]{77.14} &  \thead[c]{318.7} & \thead[c]{35.44} & \thead[c]{1909} & \thead[c]{4992} \\
 \hline 
  \thead[c]{\textsf{UAE-Q}}& {\color{black}\thead[c]{{0.5MB}}}
  &\thead[c]{2.41} & \thead[c]{1.50}&\thead[c]{7.0} & \thead[c]{36.0}& \thead[c]{17.4} & \thead[c]{1.98} & \thead[c]{34} & \thead[c]{2694} \\
 \hline\hline
 \thead[c]{Sampling}& {\color{black}\thead[c]{{0.5MB}}}&\thead[c]{2.73} & \thead[c]{1.333}&\thead[c]{12.0} &\thead[c]{49.0} & \thead[c]{2.08} & \thead[c]{1.353} & \thead[c]{6.0} & \thead[c]{41.0}\\
 \hline
 \thead[c]{BayesNet} &\thead[c]{\color{black}\thead[c]{{0.6MB}}}& \thead[c]{1.734}& \thead[c]{1.239}& \thead[c]{3.786}& \thead[c]{70.0} & \thead[c]{1.822} & \thead[c]{1.240} &  \thead[c]{4.0}& \thead[c]{98.0}\\
 \hline
 \thead[c]{KDE} &{\color{black}\thead[c]{{0.5MB}}}&\thead[c]{4.249} &\thead[c]{1.865} &\thead[c]{14.38} &\thead[c]{215.0} & \thead[c]{4.374} &\thead[c]{1.996}  & \thead[c]{14.13} &\thead[c]{177.6} \\
 \hline
 \thead[c]{DeepDB}& {\color{black}\thead[c]{{0.5MB}}} &\thead[c]{1.876} & \thead[c]{1.333}& \thead[c]{4.586}& \thead[c]{15.66}& \thead[c]{1.660} & \thead[c]{1.216} & \thead[c]{3.776} & \thead[c]{21.0} \\
 \hline
 \thead[c]{Naru}& {\color{black}\thead[c]{{0.5MB}}}
 &\thead[c]{1.775} &\thead[c]{1.258} &\thead[c]{4.0} &\thead[c]{14.5} & \thead[c]{1.925} & \thead[c]{1.324} & \thead[c]{4.565} & \thead[c]{35.0}\\
 \hline \hline
 \thead[c]{MSCN+sampling}& {\color{black}\thead[c]{{0.5MB}}}
 &\thead[c]{2.495} & \thead[c]{1.729}& \thead[c]{5.818}& \thead[c]{71.2}& \thead[c]{11.45} & \thead[c]{3.11} & \thead[c]{52.33} &\thead[c]{338.0} \\
 \hline
 \thead[c]{Feedback-KDE}& {\color{black}\thead[c]{{0.5MB}}}
 & \thead[c]{4.214} &\thead[c]{1.861} &\thead[c]{14.0} &\thead[c]{215.0} &\thead[c]{4.335}  & \thead[c]{1.962} & \thead[c]{14.02} &\thead[c]{177.6} \\
 \hline \hline
 \thead[c]{\SeSCE (Ours)}& {\color{black}\thead[c]{{0.5MB}}}
 &  \thead[c]{\textbf{1.462}} & \thead[c]{\textbf{1.196}} & \thead[c]{\textbf{3.0}} & \thead[c]{\textbf{9.0}} & \thead[c]{\textbf{1.333}} & \thead[c]{\textbf{1.138}} & \thead[c]{\textbf{2.25}} &\thead[c]{\textbf{7.0}} \\
 \hline 
\end{tabular}} \caption{Estimation Errors on Census} \label{table.3}
\end{minipage}
\vspace{-0.5em}
\end{table*}
\begin{table}[ht]
  \centering
\vspace{-2.3em}
{\color{black}
 \begin{minipage}[t]{0.49\textwidth}
  \centering
        \makeatletter\def\@captype{table}\makeatother
\scalebox{0.76}{\begin{tabular}{|l|l|l|l|l|l|l|l|l|l|}
  \hline
  \thead[c]{Model} & \thead[c]{{\color{black}Size}}&  \multicolumn{4}{c|}{\thead[c]{In-workload Queries}} &  \multicolumn{4}{c|}{\thead[c]{Random Queries}}\\
  \cline{3-10} 
   & & \thead[c]{Mean} &  \thead[c]{Median} &  \thead[c]{95th} &  \thead[c]{MAX} &  \thead[c]{Mean} &  \thead[c]{Median} &  \thead[c]{95th} &  \thead[c]{MAX} \\
  \hline
\thead[c]{LR}& {\color{black}\thead[c]{{17KB}}} & \thead[c]{3.65} & \thead[c]{2.97}& \thead[c]{14.3}&\thead[c]{314} & \thead[c]{674} & \thead[c]{406} &  \thead[c]{3461}& \thead[c]{$10^4$}\\
 \hline
 \thead[c]{MSCN-base}& \thead[c]{0.5MB} & \thead[c]{1.990} & \thead[c]{1.412}& \thead[c]{4.00}&\thead[c]{139} & \thead[c]{538} & \thead[c]{219} &  \thead[c]{2132}& \thead[c]{7231}\\
 \hline 
  \thead[c]{\textsf{UAE-Q}}& {\color{black}\thead[c]{{3.4MB}}}
  &\thead[c]{1.528} & \thead[c]{1.174}&\thead[c]{3.0} & \thead[c]{56.0}& \thead[c]{2.50} & \thead[c]{1.378} & \thead[c]{5.0} & \thead[c]{690} \\
 \hline\hline
 \thead[c]{Sampling}& {\color{black}\thead[c]{{3.4MB}}}
 & \thead[c]{3.51} & \thead[c]{1.19}& \thead[c]{17.0}&\thead[c]{99.0} & \thead[c]{2.71} & \thead[c]{\textbf{1.09}} &  \thead[c]{10.05}& \thead[c]{106}\\
 \hline
 \thead[c]{BayesNet} & \thead[c]{4.4MB} & \thead[c]{1.632} & \thead[c]{1.385}& \thead[c]{3.754}&\thead[c]{56.4} & \thead[c]{1.97} & \thead[c]{1.203} &  \thead[c]{4.10}& \thead[c]{690}\\
 \hline
 \thead[c]{KDE} & \thead[c]{3.3MB} & \thead[c]{1.519} & \thead[c]{1.161}& \thead[c]{3.000}&\thead[c]{28.67} & \thead[c]{2.145} & \thead[c]{1.166} &  \thead[c]{4.000}& \thead[c]{690}\\
 \hline
 \thead[c]{DeepDB}& \thead[c]{3.2MB} & \thead[c]{1.385} & \thead[c]{\textbf{1.10}}& \thead[c]{2.032}&\thead[c]{18.29} & \thead[c]{\textbf{1.281}} & \thead[c]{1.112} &  \thead[c]{\textbf{2.0}}& \thead[c]{\textbf{14.18}}\\
 \hline
 \thead[c]{Naru}& {\color{black}\thead[c]{{3.4MB}}}
 &\thead[c]{1.594} &\thead[c]{1.258} &\thead[c]{3.0} &\thead[c]{23.429} & \thead[c]{2.156} & \thead[c]{1.286} & \thead[c]{3.808} & \thead[c]{690}\\
 \hline \hline
 \thead[c]{MSCN+sampling}& {\color{black}\thead[c]{{3.4MB}}}
  & \thead[c]{1.717} & \thead[c]{1.257}& \thead[c]{3.882}&\thead[c]{26.0} & \thead[c]{68.43} & \thead[c]{8.61} &  \thead[c]{299}& \thead[c]{9045}\\
 \hline
 \thead[c]{Feedback-KDE}& \thead[c]{3.3MB} & \thead[c]{1.528} & \thead[c]{1.166}& \thead[c]{3.037}&\thead[c]{28.67} & \thead[c]{2.142} & \thead[c]{1.167} &  \thead[c]{4.0}& \thead[c]{690}\\
 \hline \hline
 \thead[c]{\SeSCE (Ours)}& {\color{black}\thead[c]{{3.4MB}}}
 &  \thead[c]{\textbf{1.305}} & \thead[c]{1.138} & \thead[c]{\textbf{2.0}} & \thead[c]{\textbf{13.0}} & \thead[c]{1.562} & \thead[c]{1.157} & \thead[c]{2.167} &\thead[c]{345} \\
 \hline 
\end{tabular}} \caption{{\color{black}Estimation Errors on Kddcup98}} \label{table.cup98}
\end{minipage}
}
\end{table}

\begin{table}[ht]
\vspace{-2em}
  \centering
  {\color{black}
  \begin{minipage}[t]{0.49\textwidth}
  \centering
        \makeatletter\def\@captype{table}\makeatother
\scalebox{0.9}{\begin{tabular}{|c|c|c|c|c|c|c|c|}
  \hline
  \thead[c]{Model} &   \thead[c]{Size} & \multicolumn{3}{c|}{\thead[c]{JOB-light-ranges-focused}} &  \multicolumn{3}{c|}{\thead[c]{JOB-light}}\\
  \cline{3-8} 
&  & \thead[c]{Median} &  \thead[c]{95th} &  \thead[c]{MAX} &  \thead[c]{Median} &  \thead[c]{95th} &  \thead[c]{MAX} \\
  \hline
  \thead[c]{DeepDB}& \thead[c]{4.0MB} & {\color{black}\thead[c]{2.96}}& {\color{black}\thead[c]{22.29}}&  {\color{black}\thead[c]{2435}}& {\color{black}\thead[c]{\textbf{1.26}}} & {\color{black}\thead[c]{376}} & {\color{black}\thead[c]{$10^5$}}\\
 \hline
 \thead[c]{MSCN+sampling}& \thead[c]{4.0MB} & {\color{black}\thead{1.33}} &{\color{black}\thead[c]{11.34}} &  {\color{black}\thead[c]{\textbf{32}}}& {\color{black}\thead[c]{$4\cdot10^4$}} & {\color{black}\thead[c]{$9\cdot10^7$}} & {\color{black}\thead[c]{\textbf{$9\cdot10^8$}}} \\
 \hline 
 \thead[c]{NeuroCard}& \thead[c]{4.1MB} &  \thead[c]{3.58} & \thead[c]{28.07}& \thead[c]{234} & \thead[c]{1.39} &  \thead[c]{6.53}&\thead[c]{\textbf{31.1}}\\
 \hline
 \thead[c]{\SeSCE (Ours)}& \thead[c]{4.1MB} & \thead[c]{\textbf{1.08}}& \thead[c]{\textbf{9.0}}& \thead[c]{63} & \thead[c]{1.56} &  \thead[c]{\textbf{6.29}}&\thead[c]{33.0}\\
 \hline
\end{tabular}} \caption{{\color{black}Estimation Errors on IMDB (Join Queries)}}  \label{table.join}
\end{minipage}
}
\vspace{-3.5em}
\end{table}

Tables~\ref{table.2} \textasciitilde ~\ref{table.join} show the experimental results of all models on both in-workload queries and random (out-of-workload) queries. 
The results show that \SeSCE matches or significantly exceeds the best estimator across the board, not only in terms of mean and median, but also in terms of max, which demonstrates the robustness of \SeSCE in handling tail of difficult queries. 
From these experimental results, we conclude several major findings as follows.


\noindent \textbf{(1)~\textsf{UAE-Q} outperforms other supervised methods in most cases and they are vulnerable to workload shifts.} We observe that the proposed \textsf{UAE-Q} outperforms 
\textsf{LR} and \textsf{MSCN-base} in most of the cases. 
We also observe that for all the supervised methods, the accuracy on random queries is much worse than that on in-workload queries. This indicates that supervised models 
may learn the data distribution
on the data regions the training workload focuses on, but are vulnerable to workload shifts. For example, on DMV, when moving from in-workload queries to random queries, \textsf{MSCN-base} degrades by $463\times$ in mean error and $27\times$ in max error.

\noindent \textbf{(2)~Unsupervised methods are more robust to workload shifts but still produce large error at tail.} The performance gap of unsupervised methods between in-workload and random queries is much smaller than those of supervised methods. Nevertheless, these unsupervised models still have large worst-case errors likely due to their optimization target of minimizing the average error. For instance, on \textrm{DMV} dataset, 
\textsf{Naru} produces 108 for max error. 


\noindent \textbf{(3)~DL-based methods outperform non-DL methods, especially at tail.} For supervised methods, deep learning (DL) models \textsf{MSCN-base} preforms significantly better  than traditional machine learning method \textsf{LR}. Also, for unsupervised methods, deep learning method \textsf{Naru} and \textsf{DeepDB} usually  perform better than non-DL methods (e.g., \textsf{BayesNet} ), especially in mean and max errors. These results demonstrate that DL can better capture complex data correlation than non-DL models. 

\noindent \textbf{(4)~\textsf{KDE}-based methods suffer from large domain sizes.} \textsf{KDE}-based methods (\textsf{KDE} and \textsf{Feedback-KDE}) perform poorly on two datasets with large domain sizes (\textrm{DMV}). Moreover, We find that \textsf{Feedback-KDE} can not enhance the performance of \textsf{KDE} 
significantly. It is likely because \textsf{KDE}-based methods suffer on these datasets inherently or the bandwidths computed by \textsf{Feedback-KDE} are not optimal on these datasets.

\noindent \textbf{(5)~\textsf{DeepDB} preforms relatively well on datasets with weak attribute correlation but degrades largely on datasets with strong attribute correlation.} {
\color{black}On the dataset with weaker attribute correlation (\textit{i.e.,} \textrm{Census}), \textsf{DeepDB} offers accurate estimates across all error quantiles. Nevertheless, on \textrm{DMV} that has strong attribute correlations, \textsf{DeepDB}'s performance drops quickly especially at tail, since the independence assumption in sum-product networks of \textsf{DeepDB} does not hold for this dataset.}

{ 
\color{black}
\noindent \textbf{(6)~Deep autoregressive models suffer from high dimensional data and \textsf{SPNs} might suffer from high NDVs.} 
We draw interesting conclusions in the comparison between deep autoregressive model-based methods (\textsf{Naru}, \textsf{UAE}) and SPNs-based method (\textsf{DeepDB}).
On two datasets with relatively high domain sizes (2K for \textrm{DMV}, 100\% unique for \textrm{DMV-large}), \textsf{DeepDB} may suffer at tail (\textit{e.g.,}  max error $3\cdot10^4$ on \textrm{DMV} random queries) while deep autoregressive model-based methods achieve much more stable accuracy. This is likely because that the histograms used in the leaf nodes of \textsf{DeepDB} cannot accurately model the distributions of attributes with high NDVs while deep autoregressive models can well capture them because they consider the probability of \textit{each} distinct value at the output layer.
On the contrary, on \textrm{Kddcup98} with 100 attributes, deep autoregressive models may degrade at tail (\textit{e.g.,} \textsf{Naru} makes  max error 690 on random queries. Although \textsf{UAE} improves \textsf{Naru} by $2\times$, it still makes  max error 345), but \textsf{DeepDB} can achieve very low max error.
This is likely because a higher dimensional dataset would contain more independent attributes. In this case, the autoregressive decomposition of autoregressive models might introduce noises to the model learning since it "forces" the model to learn the correlation between independent attributes. However, \textsf{DeepDB} would not have this problem as \textsf{SPNs} successfully separate those independent attributes into different groups for this dataset. The result indicates that a promising future work would be to combine the best of deep autoregressive models and \textsf{SPNs} for high-dimensional datasets with high NDVs.
}

\noindent \textbf{(7) Additional data information boosts supervised methods by a large margin.} Compared to \textsf{MSCN-base}, \textsf{MSCN+sampling} achieves much better performance on all datasets. The improvements become more obvious on random queries. The results demonstrate that including the estimates from sampling as extra features together with query features improve the accuracy of neural networks in \textsf{MSCN}. We also note that integrating query as supervised information in KDE does not help on the three datasets. 

{\color{black}
\noindent \textbf{(8) \SeSCE outperforms both of its two modules.} For example, on \textrm{DMV} both \textsf{UAE-D} and  \textsf{UAE-Q} have a max  error 108. However, \SeSCE is able to achieve max  error 5.0, which greatly improves the tail behaviour. This demonstrates the effectiveness of the unified modeling  and training in \SeSCE for cardinality estimation. }


\noindent \textbf{(9)~\SeSCE achieves the best performance on in-workload queries while maintaining the robustness on random queries.}  \SeSCE achieves the best overall results (in mean and max errors) on in-workload queries on all single-table datasets. For instance, on \textrm{DMV}, \SeSCE outperforms the second best method \textsf{Naru} by $21\times$ at tail. Also, \SeSCE produces the lowest median errors on in-workload queries on most datasets.
{\color{black} Additionally,  \SeSCE also achieves the best or comparable overall performance for join queries. As shown in Table~\ref{table.join}, on \textsf{JOB-light-ranges-focused},  \SeSCE produces the lowest median error and beats two newest data-driven models (\textsf{DeepDB}, \textsf{NeuroCard}) by a large margin across all error quantiles. Although \textsf{MSCN+sampling} outperforms \SeSCE at tail on this workload, its performance drops 
on random queries (\textsf{JOB-light}) while \SeSCE does not.
}
Surprisingly, \SeSCE can even achieve the best overall performance on random queries on \textrm{DMV} and \textrm{Census}, \textit{e.g.,} 
7.0 \textit{vs.} 21.0 in max error on \textrm{Census}, compared to the second best method \textsf{DeepDB}. It is likely because the supervised component using query workload enforces \SeSCE to offer accurate estimates on some tricky data regions (\textit{e.g.,} long tail data) that all other models cannot successfully handle. 
In addition, \SeSCE's performance on random queries on other datasets matches or outperforms that of \textsf{Naru} (or \textsf{NeuroCard}). These results demonstrate that \SeSCE can effectively leverage query workload as supervised information for enhancing the unsupervised autoregressive model, while keeping the knowledge of overall data distribution.
\vspace{-0.5em}

\subsection{Hyper-parameter Studies}
We report in-depth experimental studies to explore the best hyper-parameter settings of \SeSCE. The results on all the datasets are qualitatively similar and we report the results on \textrm{DMV} only.
\eat{
\begin{figure}[htbp]
\flushleft 
\subfigure[]{
\begin{minipage}[t]{0.5\linewidth}
\centering
\includegraphics[width=1.85in]{tau-mean}
\end{minipage}%
}%
\subfigure[]{
\begin{minipage}[t]{0.5\linewidth}
\centering
\includegraphics[width=1.85in]{tau-max}
\end{minipage}%
}%
\centering
\vspace{-2.3em}
\caption{ The effect of temperature parameter $\tau$ on \SeSCE on DMV, estimated by 2K samples.}\label{fig.tau}
\vspace{-1em}
\end{figure}
}

\noindent {\bf Impact of Temperature $\tau$} \quad
The temperature parameter $\tau$ is used in the Gumbel-Softmax algorithm which works for the supervised part of \SeSCE (\textit{i.e.,} \textsf{UAE-Q}). We have to isolate the influence from \SeSCE's unsupervised part (\textit{i.e.,} \textsf{UAE-D}) when training with \textsf{UAE-Q}. To this end, we first train \SeSCE only with \textsf{UAE-D} to obtain the overall data knowledge. Next \SeSCE is refined by \textsf{UAE-Q} with various settings of $\tau$. Specifically, 10K training queries are generated following the procedure described in Section~\ref{exp.workload} and the evaluation is conducted on 2K in-workload queries since we are interested in the effect of $\tau$ on the performance of \SeSCE's supervised part. As discussed in~\cite{jang2017categorical}, fixed $\tau$ between 0.5 and 1.0 yields good results empirically. We thus evaluate candidate values  \{0.5, 0.75, 1.0, 1.25\} for $\tau$. In addition, we use 2K samples for estimation because we are interested in the limits of \SeSCE influenced by $\tau$ and lower numbers of estimation samples also share the same trend. 
We empirically find that at $\tau$ = 1.0, \SeSCE achieves the lowest estimation errors.


\begin{figure}[htbp]
\vspace{-1.8em}
\centering
\subfigure[]{
\begin{minipage}[t]{0.5\linewidth}
\centering
\includegraphics[width=1.9in]{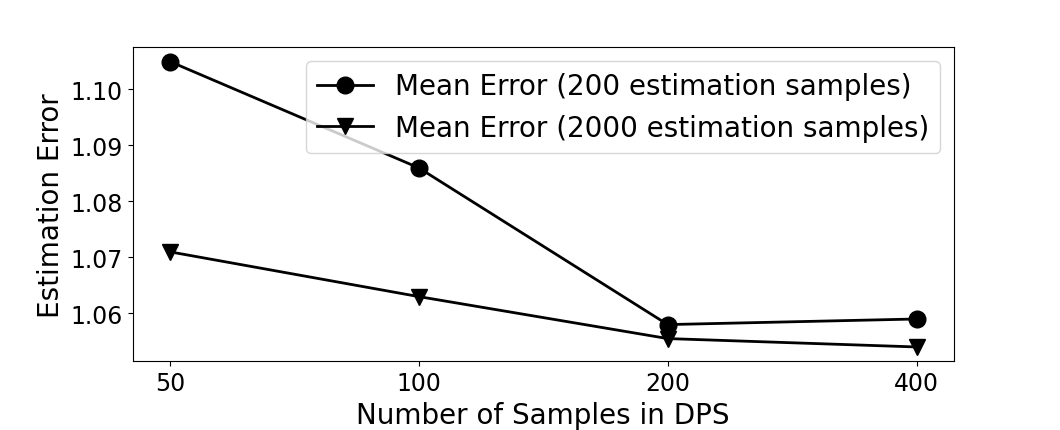}
\vspace{-2em}
\caption*{\color{black}{(a) Impact of $S$ in \textsf{DPS}}}
\end{minipage}%
 }%
\subfigure[]{
\begin{minipage}[t]{0.5\linewidth}
\centering
\includegraphics[width=1.9in]{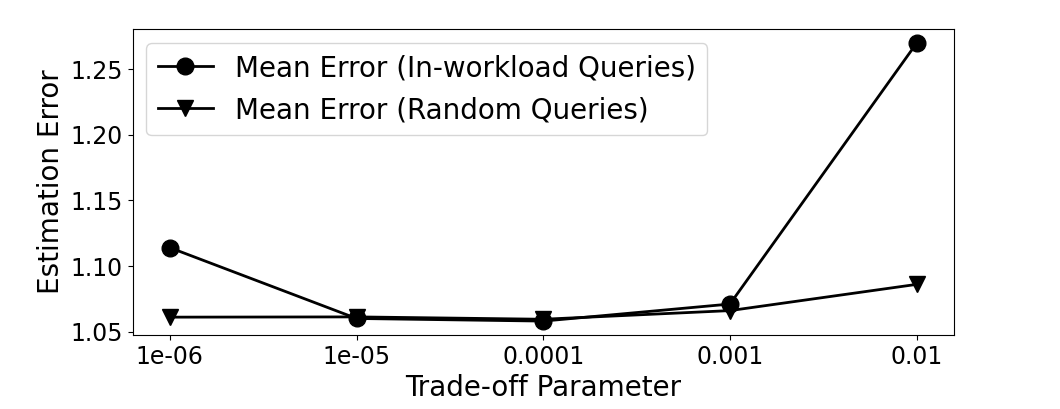}
\vspace{-2em}
\caption*{\color{black}{(b) Impact of $\lambda$}}
\end{minipage}
}%
\vspace{-1.6em}
\caption{{\color{black}The effect of hyper-parameters of \SeSCE on DMV.}} \label{fig.hyper}
\vspace{-1.4em}
\end{figure}

\noindent {\textbf{Impact of the number of training samples $S$ in \textsf{DPS}}.} 
Like $\tau$, $S$ also belongs to \textsf{UAE-Q}. Therefore, we use the same experimental setting as in $\tau$. We evaluate the values  \{50, 100, 200, 400\} for $S$, as larger numbers beyond 400 will significantly increase training overhead (i.e., consume more GPU memory). From Figure~\ref{fig.hyper} (a), 
we observe that if the number of estimation samples is set to 200, the best setting for $S$ would be 200 as well; If the number of estimation samples is increased to 2K, 400 becomes the best setting for $S$. 
%
%
%
In most cases \SeSCE can offer very accurate estimates on in-workload queries with 200 samples. Furthermore, considering the training and estimation overheads, we suggest 200 for $S$ on these datasets.


\noindent {\textbf{Impact of Trade-off Parameter $\lambda$}.} Trade-off parameter $\lambda$ rescales the losses produced by two parts of \SeSCE, \textsf{UAE-D} and \textsf{UAE-Q}. We thus use the same query workload in Section 5.1.3. The candidate values of $\lambda$ are {\color{black} \{$1e^{-6}$,$1e^{-5}$,$1e^{-4}$, $1e^{-3}$, $1e^{-2}$\}}. Figure~\ref{fig.hyper} (b) shows the performance of \SeSCE on both in-workload and random queries as $\lambda$ is varied, from which we conclude $\lambda$'s best setting is $1e^{-4}$. Moreover, when $\lambda$ is larger than $1e^{-3}$, the performance drops quickly on both kinds of queries, indicating that putting too much emphasis on \textsf{UAE-Q} will negatively affect model training and is not encouraged.
\label{subsection:63}

\vspace{-1.6em}
\subsection{\mbox{Incremental Data and Query Workload}} \label{subsection:64}

This experiment is to study the incremental learning ability of \SeSCE.
Since the ability of autoregressive models to incorporate incremental data has been demonstrated in previous work~\cite{naru, hasan2020deep}, \SeSCE, based on autoregressive model, can inherently handle incremental data. Consequently, we will not repeat the experiment in this paper. 
Beyond the previous work, we aim to show that 
\SeSCE is also able to incorporate incremental query workload while previous work~\cite{naru, hasan2020deep} cannot. 
To this end, using the same procedure in Section 5.1.3, we generate 5 parts of query workload with different query center {\color{black} for the bounded column, \textit{i.e.,} different query workload focuses on different data region}. Each part consists of 4K training queries and 200 in-workload test queries since we our goal is to demonstrate \SeSCE's effectiveness of incorporating new query workload. After training \SeSCE with the underlying data, we ingest each partition of query workload in order, following the experimental setting of incremental data in~\cite{naru}. Evaluations are conducted after each ingest using the test queries in that partition.
\begin{table}[]
    \centering
        \scalebox{0.85}{
    \begin{tabular}{|c|c|c|c|c|c|}
        \hline
        \textbf{Ingested Partitions} & \textbf{1} & \textbf{2} & \textbf{3} & \textbf{4} & \textbf{5} \\
        \hline
        \textsf{Naru}: mean & 1.035&  1.047&  1.152& 1.197 & 2.903  
        \\
        \hline
         \textsf{UAE}: mean & \textbf{1.031}& \textbf{1.039} & \textbf{1.095} & \textbf{1.132} & \textbf{1.073}  \\
        \hline       
    \end{tabular}}
    \caption{Effectiveness of incorporating incremental query workload. Stale \textsf{Naru} \textit{vs.} Refined \textsf{UAE}.}
    \label{table.increment}
    \vspace{-3em}
\end{table}
We compare refined \SeSCE to the model only trained with the underlying data (i.e., \textsf{Naru}), which cannot further ingest incremental query workload, on DMV. Table~\ref{table.increment} shows the mean errors of both methods, which are estimated by 200 samples. From the table, we observe: (1) due to the incapability of leveraging query workload, the performance of \textsf{Naru} is not stable on queries of various workloads. 
(2) \SeSCE can offer consistently accurate estimates after being refined by each query workload, which demonstrates the ability of \SeSCE in effectively ingesting  incremental query workload.
\vspace{-0.7em} 
\subsection{\mbox{Training Time $\&$ Estimation Efficiency}} \label{subsection:65}
\begin{figure}[htbp]
\vspace{-2.1em} 
\flushleft 
\subfigure[]{
~~\begin{minipage}[t]{0.5\linewidth}
\centering 
\includegraphics[width=1.85in]{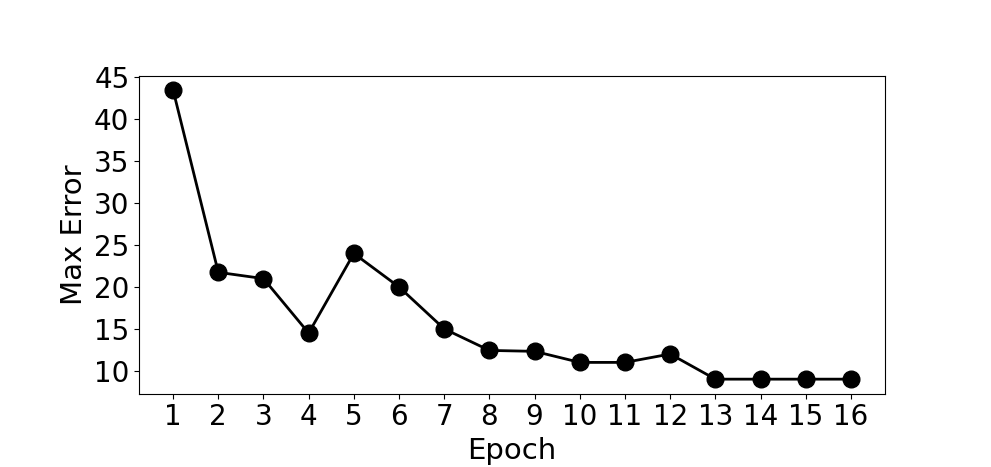}
\end{minipage}%
}%
\subfigure[]{
\begin{minipage}[t]{0.5\linewidth}
\centering
\includegraphics[height=0.85in]{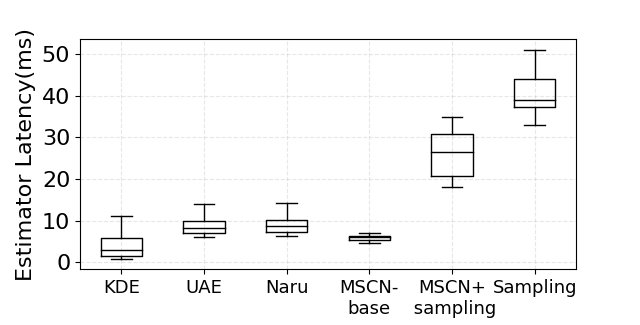}
\end{minipage}%
}%
\centering
\vspace{-3em}
\caption{(1) Left: Training epoch \textit{v.s.} max error; (2) Right: Estimation timings of different estimators.}\label{fig.timings}
\vspace{-0.5em}
\end{figure}
An epoch of \SeSCE takes about 363 seconds, 62 seconds, 
and 657 seconds on DMV, Census, 
and Kddcup98, respectively. We report the changing process of max error estimated by 200 samples as training progresses on Census in-workload queries in  Figure~\ref{fig.timings} (1). We observe  that about 13 epochs for \SeSCE yields the performance of single-digit max error, which is 9.0.

On all datasets, \SeSCE can produce estimates in around 10ms on a V100 GPU. 
 Figure~\ref{fig.timings} (2) shows the estimation latencies of different estimators on \textrm{DMV}. As shown in the figure,  \SeSCE  can produce estimates in reasonable efficiency, much faster than sampling-based methods (\textsf{MSCN+sampling}, \textsf{Sampling}). 
\vspace{-0.7em} 

\subsection{Impact on Query Optimization} \label{exp.qo}
\begin{figure}[!t]
\flushleft 
\centering
\includegraphics[width=3.35in]{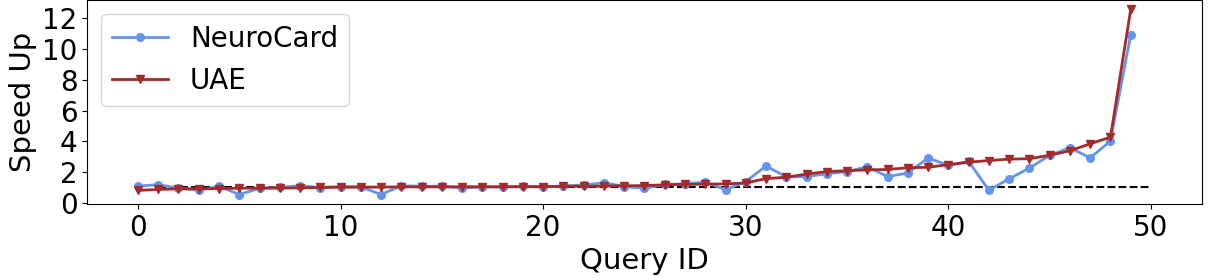}
\centering
\vspace{-2.3em}
\caption{Query execution time speed-ups.}\label{fig.qo}
\vspace{-1.6em}
\end{figure}

We proceed to evaluate the impact of \textsf{UAE} on query optimization, compared to \textsf{PostgreSQL} and \textsf{NeuroCard}. We follow the procedure~\cite{cai2019pessimistic} and we modify the source code of \textsf{PostgreSQL} to allow it accept external cardinality estimates. Then, for each query we collect the cardinality estimates of its subqueries returned by different estimators and inject them into the modified \textsf{PostgreSQL}. We use the \textsf{JOB-M}~\cite{leis2015good} benchmark as the testbed for this case because it has a more complex join schema, which is more challenging for query optimization. 
We generate 50 test queries using a template of \textsf{JOB-M} (including 6 tables and multi-way joins), following the generation procedure of \textsf{JOB-light-ranges-focused}. For training \SeSCE, we use the same template to randomly generate 10K subqueries (including 2$\sim$5 tables).
Figure~\ref{fig.qo} shows the impact of cardinality estimates from \textsf{NeuroCard} and \SeSCE on query performance compared to  \textsf{PostgreSQL}.
We have two major findings. First, more accurate cardinality estimates from deep autoregressive model-based estimators could translate into better query plans in query optimization of \textsf{PostgreSQL}. 
Second, for in-workload queries, \SeSCE could result in equivalent or better query plans to 
improve the quality of query optimization 
 without any significant slowdown compared with \textsf{PostgreSQL} and \textsf{NeuroCard} .

\section{CONCLUSIONS and future work}
We propose a novel unified deep autoregressive model that is able to utilize both data as unsupervised information and query workload as supervised information for cardinality estimation. 
%
%
%
Experiments demonstrate that \SeSCE  achieves the four goals in Section~\ref{section.1}.


We see this work as the first step toward a unified deep learning model that is able to train a single model exploiting both data information and workload information for cardinality estimation. 
 We expect that our  unified model for cardinality estimation would be inspirational to future developments of cardinality estimation models that  fuse data information and query workload. %
 We believe that our model will open interesting and promising future research directions.
For example,  exploring the power of \textsf{UAE-Q} on database generation is a very promising direction.
The generative characteristic of \textsf{UAE-Q} allows us to efficiently sample tuples from the model. This is not the case in other supervised models because it is hard to obtain the normalizing constant~\cite{chow2003probability} of the data probability for these models. This characteristic makes \textsf{UAE-Q} suitable for database generation for DBMS testing and benchmarking~\cite{arasu2011data, lo2010generating,li2018touchstone}, another important task for big data management.

\begin{acks}
This research was conducted at Singtel Cognitive and Artificial Intelligence Lab for Enterprises (SCALE@NTU), which is a collaboration between Singapore Telecommunications Limited (Singtel) and Nanyang Technological University (NTU) that is funded by the Singapore Government through the Industry Alignment Fund ‐ Industry Collaboration Projects Grant. This work was also supported in part by a MOE Tier-2 grant MOE2019-T2-2-181, a MOE Tier-1 grant RG114/19, and an NTU ACE grant. We would like to thank Zizhong Meng (NTU) for helping with  some of the experiments, and 
the anonymous reviewers for
providing constructive feedback and valuable suggestions.
\end{acks}
\bibliographystyle{ACM-Reference-Format}
\bibliography{sample-base}


\begin{thebibliography}{80}


\ifx \showCODEN    \undefined \def \showCODEN     #1{\unskip}     \fi
\ifx \showDOI      \undefined \def \showDOI       #1{#1}\fi
\ifx \showISBNx    \undefined \def \showISBNx     #1{\unskip}     \fi
\ifx \showISBNxiii \undefined \def \showISBNxiii  #1{\unskip}     \fi
\ifx \showISSN     \undefined \def \showISSN      #1{\unskip}     \fi
\ifx \showLCCN     \undefined \def \showLCCN      #1{\unskip}     \fi
\ifx \shownote     \undefined \def \shownote      #1{#1}          \fi
\ifx \showarticletitle \undefined \def \showarticletitle #1{#1}   \fi
\ifx \showURL      \undefined \def \showURL       {\relax}        \fi
\providecommand\bibfield[2]{#2}
\providecommand\bibinfo[2]{#2}
\providecommand\natexlab[1]{#1}
\providecommand\showeprint[2][]{arXiv:#2}

\bibitem[\protect\citeauthoryear{??}{dee}{[n.d.]}]%
        {deepdbcode}
 \bibinfo{year}{[n.d.]}\natexlab{}.
\newblock \bibinfo{title}{DeepDB}.
\newblock
\newblock
\newblock
\shownote{\url{https://github.com/DataManagementLab/deepdb-public/}.}


\bibitem[\protect\citeauthoryear{??}{fee}{[n.d.]}]%
        {feedbackkde}
 \bibinfo{year}{[n.d.]}\natexlab{}.
\newblock \bibinfo{title}{Feedback-KDE}.
\newblock
\newblock
\newblock
\shownote{\url{https://bitbucket.org/mheimel/feedback-kde/}.}


\bibitem[\protect\citeauthoryear{??}{msc}{[n.d.]}]%
        {mscncode}
 \bibinfo{year}{[n.d.]}\natexlab{}.
\newblock \bibinfo{title}{MSCN}.
\newblock
\newblock
\newblock
\shownote{\url{https://github.com/andreaskipf/learnedcardinalities}.}


\bibitem[\protect\citeauthoryear{??}{nar}{[n.d.]}]%
        {narucode}
 \bibinfo{year}{[n.d.]}\natexlab{}.
\newblock \bibinfo{title}{Naru-project}.
\newblock
\newblock
\newblock
\shownote{\url{https://github.com/naru-project/naru/}.}


\bibitem[\protect\citeauthoryear{??}{pos}{[n.d.]}]%
        {postgres}
 \bibinfo{year}{[n.d.]}\natexlab{}.
\newblock \bibinfo{title}{PostgreSQL}.
\newblock
\newblock
\newblock
\shownote{\url{https://www.postgresql.org/}.}


\bibitem[\protect\citeauthoryear{??}{uci}{[n.d.]}]%
        {uci}
 \bibinfo{year}{[n.d.]}\natexlab{}.
\newblock \bibinfo{title}{UCI machine learning repository}.
\newblock
\newblock
\newblock
\shownote{\url{https://archive.ics.uci.edu/ml/index.php}.}


\bibitem[\protect\citeauthoryear{??}{msd}{2017}]%
        {msdb}
 \bibinfo{year}{2017}\natexlab{}.
\newblock \bibinfo{title}{Microsoft database}.
\newblock
\newblock
\newblock
\shownote{\url{https://docs.microsoft.com/en-us/sql/relational-databases/statistics/statistics?view=sql-server-2017}.}


\bibitem[\protect\citeauthoryear{Aboulnaga and Chaudhuri}{Aboulnaga and
  Chaudhuri}{1999}]%
        {aboulnaga1999self}
\bibfield{author}{\bibinfo{person}{Ashraf Aboulnaga} {and}
  \bibinfo{person}{Surajit Chaudhuri}.} \bibinfo{year}{1999}\natexlab{}.
\newblock \showarticletitle{Self-tuning histograms: Building histograms without
  looking at data}.
\newblock \bibinfo{journal}{\emph{ACM SIGMOD Record}} \bibinfo{volume}{28},
  \bibinfo{number}{2} (\bibinfo{year}{1999}), \bibinfo{pages}{181--192}.
\newblock


\bibitem[\protect\citeauthoryear{Anagnostopoulos and
  Triantafillou}{Anagnostopoulos and Triantafillou}{2015}]%
        {anagnostopoulos2015learning}
\bibfield{author}{\bibinfo{person}{Christos Anagnostopoulos} {and}
  \bibinfo{person}{Peter Triantafillou}.} \bibinfo{year}{2015}\natexlab{}.
\newblock \showarticletitle{Learning to accurately count with query-driven
  predictive analytics}. In \bibinfo{booktitle}{\emph{2015 IEEE international
  conference on big data (big data)}}. IEEE, \bibinfo{pages}{14--23}.
\newblock


\bibitem[\protect\citeauthoryear{Arasu, Kaushik, and Li}{Arasu
  et~al\mbox{.}}{2011}]%
        {arasu2011data}
\bibfield{author}{\bibinfo{person}{Arvind Arasu}, \bibinfo{person}{Raghav
  Kaushik}, {and} \bibinfo{person}{Jian Li}.} \bibinfo{year}{2011}\natexlab{}.
\newblock \showarticletitle{Data generation using declarative constraints}. In
  \bibinfo{booktitle}{\emph{SIGMOD}}. \bibinfo{pages}{685--696}.
\newblock


\bibitem[\protect\citeauthoryear{Bottou}{Bottou}{2010}]%
        {bottou2010large}
\bibfield{author}{\bibinfo{person}{L{\'e}on Bottou}.}
  \bibinfo{year}{2010}\natexlab{}.
\newblock \showarticletitle{Large-scale machine learning with stochastic
  gradient descent}.
\newblock In \bibinfo{booktitle}{\emph{Proceedings of COMPSTAT'2010}}.
  \bibinfo{publisher}{Springer}, \bibinfo{pages}{177--186}.
\newblock


\bibitem[\protect\citeauthoryear{Bruno, Chaudhuri, and Gravano}{Bruno
  et~al\mbox{.}}{2001}]%
        {bruno2001stholes}
\bibfield{author}{\bibinfo{person}{Nicolas Bruno}, \bibinfo{person}{Surajit
  Chaudhuri}, {and} \bibinfo{person}{Luis Gravano}.}
  \bibinfo{year}{2001}\natexlab{}.
\newblock \showarticletitle{STHoles: a multidimensional workload-aware
  histogram}. In \bibinfo{booktitle}{\emph{SIGMOD}}. \bibinfo{pages}{211--222}.
\newblock


\bibitem[\protect\citeauthoryear{Cai, Balazinska, and Suciu}{Cai
  et~al\mbox{.}}{2019}]%
        {cai2019pessimistic}
\bibfield{author}{\bibinfo{person}{Walter Cai}, \bibinfo{person}{Magdalena
  Balazinska}, {and} \bibinfo{person}{Dan Suciu}.}
  \bibinfo{year}{2019}\natexlab{}.
\newblock \showarticletitle{Pessimistic cardinality estimation: Tighter upper
  bounds for intermediate join cardinalities}. In
  \bibinfo{booktitle}{\emph{SIGMOD}}. \bibinfo{pages}{18--35}.
\newblock


\bibitem[\protect\citeauthoryear{Chow and Liu}{Chow and Liu}{1968}]%
        {chow1968approximating}
\bibfield{author}{\bibinfo{person}{C Chow} {and} \bibinfo{person}{Cong Liu}.}
  \bibinfo{year}{1968}\natexlab{}.
\newblock \showarticletitle{Approximating discrete probability distributions
  with dependence trees}.
\newblock \bibinfo{journal}{\emph{IEEE transactions on Information Theory}}
  \bibinfo{volume}{14}, \bibinfo{number}{3} (\bibinfo{year}{1968}),
  \bibinfo{pages}{462--467}.
\newblock


\bibitem[\protect\citeauthoryear{Chow and Teicher}{Chow and Teicher}{2003}]%
        {chow2003probability}
\bibfield{author}{\bibinfo{person}{Yuan~Shih Chow} {and} \bibinfo{person}{Henry
  Teicher}.} \bibinfo{year}{2003}\natexlab{}.
\newblock \bibinfo{booktitle}{\emph{Probability theory: independence,
  interchangeability, martingales}}.
\newblock \bibinfo{publisher}{Springer Science \& Business Media}.
\newblock


\bibitem[\protect\citeauthoryear{Cormode, Garofalakis, Haas, and
  Jermaine}{Cormode et~al\mbox{.}}{2012}]%
        {survey2012}
\bibfield{author}{\bibinfo{person}{Graham Cormode}, \bibinfo{person}{Minos
  Garofalakis}, \bibinfo{person}{Peter~J. Haas}, {and} \bibinfo{person}{Chris
  Jermaine}.} \bibinfo{year}{2012}\natexlab{}.
\newblock \showarticletitle{Synopses for Massive Data: Samples, Histograms,
  Wavelets, Sketches}.
\newblock \bibinfo{journal}{\emph{Found. Trends Databases}}
  \bibinfo{volume}{4}, \bibinfo{number}{1–3} (\bibinfo{date}{Jan.}
  \bibinfo{year}{2012}), \bibinfo{pages}{1–294}.
\newblock
\showISSN{1931-7883}
\urldef\tempurl%
\url{https://doi.org/10.1561/1900000004}
\showDOI{\tempurl}


\bibitem[\protect\citeauthoryear{Deshpande, Garofalakis, and Rastogi}{Deshpande
  et~al\mbox{.}}{2001}]%
        {deshpande2001independence}
\bibfield{author}{\bibinfo{person}{Amol Deshpande}, \bibinfo{person}{Minos
  Garofalakis}, {and} \bibinfo{person}{Rajeev Rastogi}.}
  \bibinfo{year}{2001}\natexlab{}.
\newblock \showarticletitle{Independence is good: Dependency-based histogram
  synopses for high-dimensional data}.
\newblock \bibinfo{journal}{\emph{ACM SIGMOD Record}} \bibinfo{volume}{30},
  \bibinfo{number}{2} (\bibinfo{year}{2001}), \bibinfo{pages}{199--210}.
\newblock


\bibitem[\protect\citeauthoryear{Doane and Seward}{Doane and Seward}{2011}]%
        {doane2011measuring}
\bibfield{author}{\bibinfo{person}{David~P Doane} {and} \bibinfo{person}{Lori~E
  Seward}.} \bibinfo{year}{2011}\natexlab{}.
\newblock \showarticletitle{Measuring skewness: a forgotten statistic?}
\newblock \bibinfo{journal}{\emph{Journal of statistics education}}
  \bibinfo{volume}{19}, \bibinfo{number}{2} (\bibinfo{year}{2011}).
\newblock


\bibitem[\protect\citeauthoryear{Dutt, Wang, Nazi, Kandula, Narasayya, and
  Chaudhuri}{Dutt et~al\mbox{.}}{2019}]%
        {dutt2019selectivity}
\bibfield{author}{\bibinfo{person}{Anshuman Dutt}, \bibinfo{person}{Chi Wang},
  \bibinfo{person}{Azade Nazi}, \bibinfo{person}{Srikanth Kandula},
  \bibinfo{person}{Vivek Narasayya}, {and} \bibinfo{person}{Surajit
  Chaudhuri}.} \bibinfo{year}{2019}\natexlab{}.
\newblock \showarticletitle{Selectivity estimation for range predicates using
  lightweight models}.
\newblock \bibinfo{journal}{\emph{VLDB}} \bibinfo{volume}{12},
  \bibinfo{number}{9} (\bibinfo{year}{2019}), \bibinfo{pages}{1044--1057}.
\newblock


\bibitem[\protect\citeauthoryear{Germain, Gregor, Murray, and
  Larochelle}{Germain et~al\mbox{.}}{2015a}]%
        {germain2015made}
\bibfield{author}{\bibinfo{person}{Mathieu Germain}, \bibinfo{person}{Karol
  Gregor}, \bibinfo{person}{Iain Murray}, {and} \bibinfo{person}{Hugo
  Larochelle}.} \bibinfo{year}{2015}\natexlab{a}.
\newblock \showarticletitle{Made: Masked autoencoder for distribution
  estimation}. In \bibinfo{booktitle}{\emph{ICML}}. \bibinfo{pages}{881--889}.
\newblock


\bibitem[\protect\citeauthoryear{Germain, Gregor, Murray, and
  Larochelle}{Germain et~al\mbox{.}}{2015b}]%
        {made}
\bibfield{author}{\bibinfo{person}{Mathieu Germain}, \bibinfo{person}{Karol
  Gregor}, \bibinfo{person}{Iain Murray}, {and} \bibinfo{person}{Hugo
  Larochelle}.} \bibinfo{year}{2015}\natexlab{b}.
\newblock \showarticletitle{MADE: Masked Autoencoder for Distribution
  Estimation}.
\newblock


\bibitem[\protect\citeauthoryear{Getoor, Taskar, and Koller}{Getoor
  et~al\mbox{.}}{2001}]%
        {getoor2001selectivity}
\bibfield{author}{\bibinfo{person}{Lise Getoor}, \bibinfo{person}{Benjamin
  Taskar}, {and} \bibinfo{person}{Daphne Koller}.}
  \bibinfo{year}{2001}\natexlab{}.
\newblock \showarticletitle{Selectivity estimation using probabilistic models}.
  In \bibinfo{booktitle}{\emph{SIGMOD}}. \bibinfo{pages}{461--472}.
\newblock


\bibitem[\protect\citeauthoryear{Goodfellow, Bengio, and Courville}{Goodfellow
  et~al\mbox{.}}{2016}]%
        {goodfellow2016deep}
\bibfield{author}{\bibinfo{person}{Ian Goodfellow}, \bibinfo{person}{Yoshua
  Bengio}, {and} \bibinfo{person}{Aaron Courville}.}
  \bibinfo{year}{2016}\natexlab{}.
\newblock \bibinfo{booktitle}{\emph{Deep learning}}.
\newblock \bibinfo{publisher}{MIT press}.
\newblock


\bibitem[\protect\citeauthoryear{Gu, Levine, Sutskever, and Mnih}{Gu
  et~al\mbox{.}}{2016}]%
        {gu2016muprop}
\bibfield{author}{\bibinfo{person}{Shixiang Gu}, \bibinfo{person}{Sergey
  Levine}, \bibinfo{person}{Ilya Sutskever}, {and} \bibinfo{person}{Andriy
  Mnih}.} \bibinfo{year}{2016}\natexlab{}.
\newblock \showarticletitle{Muprop: Unbiased backpropagation for stochastic
  neural networks}. In \bibinfo{booktitle}{\emph{ICLR}}.
\newblock


\bibitem[\protect\citeauthoryear{Gunopulos, Kollios, Tsotras, and
  Domeniconi}{Gunopulos et~al\mbox{.}}{2000}]%
        {gunopulos2000approximating}
\bibfield{author}{\bibinfo{person}{Dimitrios Gunopulos},
  \bibinfo{person}{George Kollios}, \bibinfo{person}{Vassilis~J Tsotras}, {and}
  \bibinfo{person}{Carlotta Domeniconi}.} \bibinfo{year}{2000}\natexlab{}.
\newblock \showarticletitle{Approximating multi-dimensional aggregate range
  queries over real attributes}.
\newblock \bibinfo{journal}{\emph{Acm Sigmod Record}} \bibinfo{volume}{29},
  \bibinfo{number}{2} (\bibinfo{year}{2000}), \bibinfo{pages}{463--474}.
\newblock


\bibitem[\protect\citeauthoryear{Gunopulos, Kollios, Tsotras, and
  Domeniconi}{Gunopulos et~al\mbox{.}}{2005}]%
        {gunopulos2005selectivity}
\bibfield{author}{\bibinfo{person}{Dimitrios Gunopulos},
  \bibinfo{person}{George Kollios}, \bibinfo{person}{Vassilis~J Tsotras}, {and}
  \bibinfo{person}{Carlotta Domeniconi}.} \bibinfo{year}{2005}\natexlab{}.
\newblock \showarticletitle{Selectivity estimators for multidimensional range
  queries over real attributes}.
\newblock \bibinfo{journal}{\emph{The VLDB Journal}} \bibinfo{volume}{14},
  \bibinfo{number}{2} (\bibinfo{year}{2005}), \bibinfo{pages}{137--154}.
\newblock


\bibitem[\protect\citeauthoryear{Haas, Naughton, and Swami}{Haas
  et~al\mbox{.}}{1994}]%
        {haas1994relative}
\bibfield{author}{\bibinfo{person}{Peter~J Haas}, \bibinfo{person}{Jeffrey~F
  Naughton}, {and} \bibinfo{person}{Arun~N Swami}.}
  \bibinfo{year}{1994}\natexlab{}.
\newblock \showarticletitle{On the relative cost of sampling for join
  selectivity estimation}. In \bibinfo{booktitle}{\emph{PODS}}.
  \bibinfo{pages}{14--24}.
\newblock


\bibitem[\protect\citeauthoryear{Hasan, Thirumuruganathan, Augustine, Koudas,
  and Das}{Hasan et~al\mbox{.}}{2020}]%
        {hasan2020deep}
\bibfield{author}{\bibinfo{person}{Shohedul Hasan}, \bibinfo{person}{Saravanan
  Thirumuruganathan}, \bibinfo{person}{Jees Augustine}, \bibinfo{person}{Nick
  Koudas}, {and} \bibinfo{person}{Gautam Das}.}
  \bibinfo{year}{2020}\natexlab{}.
\newblock \showarticletitle{Deep Learning Models for Selectivity Estimation of
  Multi-Attribute Queries}. In \bibinfo{booktitle}{\emph{SIGMOD}}.
  \bibinfo{pages}{1035--1050}.
\newblock


\bibitem[\protect\citeauthoryear{Hayek and Shmueli}{Hayek and Shmueli}{2020}]%
        {hayek2019improved}
\bibfield{author}{\bibinfo{person}{Rojeh Hayek} {and} \bibinfo{person}{Oded
  Shmueli}.} \bibinfo{year}{2020}\natexlab{}.
\newblock \showarticletitle{Improved Cardinality Estimation by Learning Queries
  Containment Rates}. In \bibinfo{booktitle}{\emph{EDBT}}.
\newblock


\bibitem[\protect\citeauthoryear{Heimel, Kiefer, and Markl}{Heimel
  et~al\mbox{.}}{2015}]%
        {heimel2015self}
\bibfield{author}{\bibinfo{person}{Max Heimel}, \bibinfo{person}{Martin
  Kiefer}, {and} \bibinfo{person}{Volker Markl}.}
  \bibinfo{year}{2015}\natexlab{}.
\newblock \showarticletitle{Self-tuning, gpu-accelerated kernel density models
  for multidimensional selectivity estimation}. In
  \bibinfo{booktitle}{\emph{SIGMOD}}. \bibinfo{pages}{1477--1492}.
\newblock


\bibitem[\protect\citeauthoryear{Hilprecht, Schmidt, Kulessa, Molina, Kersting,
  and Binnig}{Hilprecht et~al\mbox{.}}{2020}]%
        {deepdb}
\bibfield{author}{\bibinfo{person}{Benjamin Hilprecht},
  \bibinfo{person}{Andreas Schmidt}, \bibinfo{person}{Moritz Kulessa},
  \bibinfo{person}{Alejandro Molina}, \bibinfo{person}{Kristian Kersting},
  {and} \bibinfo{person}{Carsten Binnig}.} \bibinfo{year}{2020}\natexlab{}.
\newblock \showarticletitle{DeepDB: Learn from Data, not from Queries!}
\newblock \bibinfo{journal}{\emph{VLDB}} \bibinfo{volume}{13},
  \bibinfo{number}{7}, \bibinfo{pages}{992--1005}.
\newblock


\bibitem[\protect\citeauthoryear{Hu, Wang, Fan, and Agarwal}{Hu
  et~al\mbox{.}}{2018}]%
        {hucost}
\bibfield{author}{\bibinfo{person}{R Hu}, \bibinfo{person}{Zhenhua Wang},
  \bibinfo{person}{W Fan}, {and} \bibinfo{person}{S Agarwal}.}
  \bibinfo{year}{2018}\natexlab{}.
\newblock \bibinfo{title}{Cost based optimizer in apache spark 2.2}.
\newblock
\newblock
\newblock
\shownote{\url{https://databricks.com/blog/2017/08/31/cost-based-optimizer-in-apache-spark-2-2.html}.}


\bibitem[\protect\citeauthoryear{Huang, Yoon, Pettie, and Mozafari}{Huang
  et~al\mbox{.}}{2019}]%
        {huang2019joins}
\bibfield{author}{\bibinfo{person}{Dawei Huang}, \bibinfo{person}{Dong~Young
  Yoon}, \bibinfo{person}{Seth Pettie}, {and} \bibinfo{person}{Barzan
  Mozafari}.} \bibinfo{year}{2019}\natexlab{}.
\newblock \showarticletitle{Joins on samples: A theoretical guide for
  practitioners}.
\newblock \bibinfo{journal}{\emph{VLDB}} (\bibinfo{year}{2019}).
\newblock


\bibitem[\protect\citeauthoryear{Ilyas, Markl, Haas, Brown, and
  Aboulnaga}{Ilyas et~al\mbox{.}}{2004}]%
        {ilyas2004cords}
\bibfield{author}{\bibinfo{person}{Ihab~F Ilyas}, \bibinfo{person}{Volker
  Markl}, \bibinfo{person}{Peter Haas}, \bibinfo{person}{Paul Brown}, {and}
  \bibinfo{person}{Ashraf Aboulnaga}.} \bibinfo{year}{2004}\natexlab{}.
\newblock \showarticletitle{CORDS: automatic discovery of correlations and soft
  functional dependencies}. In \bibinfo{booktitle}{\emph{SIGMOD}}.
  \bibinfo{pages}{647--658}.
\newblock


\bibitem[\protect\citeauthoryear{Jagadish, Jin, Ooi, and Tan}{Jagadish
  et~al\mbox{.}}{2001}]%
        {jagadish2001global}
\bibfield{author}{\bibinfo{person}{HV Jagadish}, \bibinfo{person}{Hui Jin},
  \bibinfo{person}{Beng~Chin Ooi}, {and} \bibinfo{person}{Kian-Lee Tan}.}
  \bibinfo{year}{2001}\natexlab{}.
\newblock \showarticletitle{Global optimization of histograms}.
\newblock \bibinfo{journal}{\emph{ACM SIGMOD Record}} \bibinfo{volume}{30},
  \bibinfo{number}{2} (\bibinfo{year}{2001}), \bibinfo{pages}{223--234}.
\newblock


\bibitem[\protect\citeauthoryear{Jang, Gu, and Poole}{Jang
  et~al\mbox{.}}{2017}]%
        {jang2017categorical}
\bibfield{author}{\bibinfo{person}{Eric Jang}, \bibinfo{person}{Shixiang Gu},
  {and} \bibinfo{person}{Ben Poole}.} \bibinfo{year}{2017}\natexlab{}.
\newblock \showarticletitle{Categorical reparameterization with
  gumbel-softmax}. In \bibinfo{booktitle}{\emph{ICLR}}.
\newblock


\bibitem[\protect\citeauthoryear{Kiefer, Heimel, Bre{\ss}, and Markl}{Kiefer
  et~al\mbox{.}}{2017}]%
        {kiefer2017estimating}
\bibfield{author}{\bibinfo{person}{Martin Kiefer}, \bibinfo{person}{Max
  Heimel}, \bibinfo{person}{Sebastian Bre{\ss}}, {and} \bibinfo{person}{Volker
  Markl}.} \bibinfo{year}{2017}\natexlab{}.
\newblock \showarticletitle{Estimating join selectivities using
  bandwidth-optimized kernel density models}.
\newblock \bibinfo{journal}{\emph{VLDB}} \bibinfo{volume}{10},
  \bibinfo{number}{13} (\bibinfo{year}{2017}), \bibinfo{pages}{2085--2096}.
\newblock


\bibitem[\protect\citeauthoryear{Kingma and Welling}{Kingma and
  Welling}{2014}]%
        {kingma2014auto}
\bibfield{author}{\bibinfo{person}{Diederik~P Kingma} {and}
  \bibinfo{person}{Max Welling}.} \bibinfo{year}{2014}\natexlab{}.
\newblock \showarticletitle{Auto-encoding variational bayes}. In
  \bibinfo{booktitle}{\emph{ICLR}}.
\newblock


\bibitem[\protect\citeauthoryear{Kipf, Kipf, Radke, Leis, Boncz, and
  Kemper}{Kipf et~al\mbox{.}}{2019}]%
        {kipf2019learned}
\bibfield{author}{\bibinfo{person}{Andreas Kipf}, \bibinfo{person}{Thomas
  Kipf}, \bibinfo{person}{Bernhard Radke}, \bibinfo{person}{Viktor Leis},
  \bibinfo{person}{Peter Boncz}, {and} \bibinfo{person}{Alfons Kemper}.}
  \bibinfo{year}{2019}\natexlab{}.
\newblock \showarticletitle{Learned cardinalities: Estimating correlated joins
  with deep learning}. In \bibinfo{booktitle}{\emph{CIDR}}.
\newblock


\bibitem[\protect\citeauthoryear{Kutner, Nachtsheim, Neter, Li,
  et~al\mbox{.}}{Kutner et~al\mbox{.}}{2005}]%
        {kutner2005applied}
\bibfield{author}{\bibinfo{person}{Michael~H Kutner},
  \bibinfo{person}{Christopher~J Nachtsheim}, \bibinfo{person}{John Neter},
  \bibinfo{person}{William Li}, {et~al\mbox{.}}}
  \bibinfo{year}{2005}\natexlab{}.
\newblock \bibinfo{booktitle}{\emph{Applied linear statistical models}}.
  Vol.~\bibinfo{volume}{5}.
\newblock \bibinfo{publisher}{McGraw-Hill Irwin New York}.
\newblock


\bibitem[\protect\citeauthoryear{Leis, Gubichev, Mirchev, Boncz, Kemper, and
  Neumann}{Leis et~al\mbox{.}}{2015}]%
        {leis2015good}
\bibfield{author}{\bibinfo{person}{Viktor Leis}, \bibinfo{person}{Andrey
  Gubichev}, \bibinfo{person}{Atanas Mirchev}, \bibinfo{person}{Peter Boncz},
  \bibinfo{person}{Alfons Kemper}, {and} \bibinfo{person}{Thomas Neumann}.}
  \bibinfo{year}{2015}\natexlab{}.
\newblock \showarticletitle{How good are query optimizers, really?}
\newblock \bibinfo{journal}{\emph{Proceedings of the VLDB Endowment}}
  \bibinfo{volume}{9}, \bibinfo{number}{3} (\bibinfo{year}{2015}),
  \bibinfo{pages}{204--215}.
\newblock


\bibitem[\protect\citeauthoryear{Leis, Radke, Gubichev, Kemper, and
  Neumann}{Leis et~al\mbox{.}}{2017}]%
        {leis2017cardinality}
\bibfield{author}{\bibinfo{person}{Viktor Leis}, \bibinfo{person}{Bernhard
  Radke}, \bibinfo{person}{Andrey Gubichev}, \bibinfo{person}{Alfons Kemper},
  {and} \bibinfo{person}{Thomas Neumann}.} \bibinfo{year}{2017}\natexlab{}.
\newblock \showarticletitle{Cardinality Estimation Done Right: Index-Based Join
  Sampling.}. In \bibinfo{booktitle}{\emph{CIDR}}.
\newblock


\bibitem[\protect\citeauthoryear{Leis, Radke, Gubichev, Mirchev, Boncz, Kemper,
  and Neumann}{Leis et~al\mbox{.}}{2018}]%
        {leis2018query}
\bibfield{author}{\bibinfo{person}{Viktor Leis}, \bibinfo{person}{Bernhard
  Radke}, \bibinfo{person}{Andrey Gubichev}, \bibinfo{person}{Atanas Mirchev},
  \bibinfo{person}{Peter Boncz}, \bibinfo{person}{Alfons Kemper}, {and}
  \bibinfo{person}{Thomas Neumann}.} \bibinfo{year}{2018}\natexlab{}.
\newblock \showarticletitle{Query optimization through the looking glass, and
  what we found running the Join Order Benchmark}.
\newblock \bibinfo{journal}{\emph{The VLDB Journal}} \bibinfo{volume}{27},
  \bibinfo{number}{5} (\bibinfo{year}{2018}), \bibinfo{pages}{643--668}.
\newblock


\bibitem[\protect\citeauthoryear{Li, Zhang, Yang, Zhang, and Zhou}{Li
  et~al\mbox{.}}{2018}]%
        {li2018touchstone}
\bibfield{author}{\bibinfo{person}{Yuming Li}, \bibinfo{person}{Rong Zhang},
  \bibinfo{person}{Xiaoyan Yang}, \bibinfo{person}{Zhenjie Zhang}, {and}
  \bibinfo{person}{Aoying Zhou}.} \bibinfo{year}{2018}\natexlab{}.
\newblock \showarticletitle{Touchstone: generating enormous query-aware test
  databases}. In \bibinfo{booktitle}{\emph{2018 $\{$USENIX$\}$ Annual Technical
  Conference ($\{$USENIX$\}$$\{$ATC$\}$ 18)}}. \bibinfo{pages}{575--586}.
\newblock


\bibitem[\protect\citeauthoryear{Liang, Yang, Stoica, Abbeel, Duan, and
  Chen}{Liang et~al\mbox{.}}{2020}]%
        {liang2020variable}
\bibfield{author}{\bibinfo{person}{Eric Liang}, \bibinfo{person}{Zongheng
  Yang}, \bibinfo{person}{Ion Stoica}, \bibinfo{person}{Pieter Abbeel},
  \bibinfo{person}{Yan Duan}, {and} \bibinfo{person}{Xi Chen}.}
  \bibinfo{year}{2020}\natexlab{}.
\newblock \showarticletitle{Variable Skipping for Autoregressive Range Density
  Estimation}.
\newblock \bibinfo{journal}{\emph{ICML}}.
\newblock


\bibitem[\protect\citeauthoryear{Lim, Wang, and Vitter}{Lim
  et~al\mbox{.}}{2003}]%
        {lim2003sash}
\bibfield{author}{\bibinfo{person}{Lipyeow Lim}, \bibinfo{person}{Min Wang},
  {and} \bibinfo{person}{Jeffrey~Scott Vitter}.}
  \bibinfo{year}{2003}\natexlab{}.
\newblock \showarticletitle{SASH: A self-adaptive histogram set for dynamically
  changing workloads}. In \bibinfo{booktitle}{\emph{Proceedings 2003 VLDB
  Conference}}. Elsevier, \bibinfo{pages}{369--380}.
\newblock


\bibitem[\protect\citeauthoryear{Lipton, Naughton, and Schneider}{Lipton
  et~al\mbox{.}}{1990}]%
        {lipton1990practical}
\bibfield{author}{\bibinfo{person}{Richard~J Lipton},
  \bibinfo{person}{Jeffrey~F Naughton}, {and} \bibinfo{person}{Donovan~A
  Schneider}.} \bibinfo{year}{1990}\natexlab{}.
\newblock \showarticletitle{Practical selectivity estimation through adaptive
  sampling}. In \bibinfo{booktitle}{\emph{SIGMOD}}. \bibinfo{pages}{1--11}.
\newblock


\bibitem[\protect\citeauthoryear{Lo, Cheng, and Hon}{Lo et~al\mbox{.}}{2010}]%
        {lo2010generating}
\bibfield{author}{\bibinfo{person}{Eric Lo}, \bibinfo{person}{Nick Cheng},
  {and} \bibinfo{person}{Wing-Kai Hon}.} \bibinfo{year}{2010}\natexlab{}.
\newblock \showarticletitle{Generating databases for query workloads}.
\newblock \bibinfo{journal}{\emph{VLDB}} \bibinfo{volume}{3},
  \bibinfo{number}{1-2} (\bibinfo{year}{2010}), \bibinfo{pages}{848--859}.
\newblock


\bibitem[\protect\citeauthoryear{Lynch}{Lynch}{1988}]%
        {lynch1988selectivity}
\bibfield{author}{\bibinfo{person}{Clifford~A Lynch}.}
  \bibinfo{year}{1988}\natexlab{}.
\newblock \showarticletitle{Selectivity Estimation and Query Optimization in
  Large Databases with Highly Skewed Distribution of Column Values.}. In
  \bibinfo{booktitle}{\emph{VLDB}}. \bibinfo{pages}{240--251}.
\newblock


\bibitem[\protect\citeauthoryear{Maddison, Mnih, and Teh}{Maddison
  et~al\mbox{.}}{2017}]%
        {maddison2017concrete}
\bibfield{author}{\bibinfo{person}{Chris~J Maddison}, \bibinfo{person}{Andriy
  Mnih}, {and} \bibinfo{person}{Yee~Whye Teh}.}
  \bibinfo{year}{2017}\natexlab{}.
\newblock \showarticletitle{The concrete distribution: A continuous relaxation
  of discrete random variables}. In \bibinfo{booktitle}{\emph{ICLR}}.
\newblock


\bibitem[\protect\citeauthoryear{Moerkotte, Neumann, and Steidl}{Moerkotte
  et~al\mbox{.}}{2009}]%
        {moerkotte2009preventing}
\bibfield{author}{\bibinfo{person}{Guido Moerkotte}, \bibinfo{person}{Thomas
  Neumann}, {and} \bibinfo{person}{Gabriele Steidl}.}
  \bibinfo{year}{2009}\natexlab{}.
\newblock \showarticletitle{Preventing bad plans by bounding the impact of
  cardinality estimation errors}.
\newblock \bibinfo{journal}{\emph{PVLDB}} \bibinfo{volume}{2},
  \bibinfo{number}{1} (\bibinfo{year}{2009}), \bibinfo{pages}{982--993}.
\newblock


\bibitem[\protect\citeauthoryear{Muralikrishna and DeWitt}{Muralikrishna and
  DeWitt}{1988}]%
        {muralikrishna1988equi}
\bibfield{author}{\bibinfo{person}{M Muralikrishna} {and}
  \bibinfo{person}{David~J DeWitt}.} \bibinfo{year}{1988}\natexlab{}.
\newblock \showarticletitle{Equi-depth multidimensional histograms}. In
  \bibinfo{booktitle}{\emph{SIGMOD}}. \bibinfo{pages}{28--36}.
\newblock


\bibitem[\protect\citeauthoryear{Nash and Durkan}{Nash and Durkan}{2019}]%
        {nash2019autoregressive}
\bibfield{author}{\bibinfo{person}{Charlie Nash} {and} \bibinfo{person}{Conor
  Durkan}.} \bibinfo{year}{2019}\natexlab{}.
\newblock \showarticletitle{Autoregressive energy machines}.
\newblock \bibinfo{journal}{\emph{ICML}}.
\newblock


\bibitem[\protect\citeauthoryear{Ortiz, Balazinska, Gehrke, and Keerthi}{Ortiz
  et~al\mbox{.}}{2019}]%
        {ortiz2019empirical}
\bibfield{author}{\bibinfo{person}{Jennifer Ortiz}, \bibinfo{person}{Magdalena
  Balazinska}, \bibinfo{person}{Johannes Gehrke}, {and}
  \bibinfo{person}{S~Sathiya Keerthi}.} \bibinfo{year}{2019}\natexlab{}.
\newblock \showarticletitle{An Empirical Analysis of Deep Learning for
  Cardinality Estimation}.
\newblock \bibinfo{journal}{\emph{arXiv preprint arXiv:1905.06425}}
  (\bibinfo{year}{2019}).
\newblock


\bibitem[\protect\citeauthoryear{Papamakarios, Pavlakou, and
  Murray}{Papamakarios et~al\mbox{.}}{2017}]%
        {papamakarios2017masked}
\bibfield{author}{\bibinfo{person}{George Papamakarios}, \bibinfo{person}{Theo
  Pavlakou}, {and} \bibinfo{person}{Iain Murray}.}
  \bibinfo{year}{2017}\natexlab{}.
\newblock \showarticletitle{Masked autoregressive flow for density estimation}.
  In \bibinfo{booktitle}{\emph{NIPS}}. \bibinfo{pages}{2338--2347}.
\newblock


\bibitem[\protect\citeauthoryear{Park, Zhong, and Mozafari}{Park
  et~al\mbox{.}}{2020}]%
        {park2020quicksel}
\bibfield{author}{\bibinfo{person}{Yongjoo Park}, \bibinfo{person}{Shucheng
  Zhong}, {and} \bibinfo{person}{Barzan Mozafari}.}
  \bibinfo{year}{2020}\natexlab{}.
\newblock \showarticletitle{Quicksel: Quick selectivity learning with mixture
  models}. In \bibinfo{booktitle}{\emph{SIGMOD}}. \bibinfo{pages}{1017--1033}.
\newblock


\bibitem[\protect\citeauthoryear{Poon and Domingos}{Poon and Domingos}{2011}]%
        {poon2011sum}
\bibfield{author}{\bibinfo{person}{Hoifung Poon} {and} \bibinfo{person}{Pedro
  Domingos}.} \bibinfo{year}{2011}\natexlab{}.
\newblock \showarticletitle{Sum-product networks: A new deep architecture}. In
  \bibinfo{booktitle}{\emph{ICCV Workshops}}. IEEE, \bibinfo{pages}{689--690}.
\newblock


\bibitem[\protect\citeauthoryear{Poosala, Haas, Ioannidis, and Shekita}{Poosala
  et~al\mbox{.}}{1996}]%
        {poosala1996improved}
\bibfield{author}{\bibinfo{person}{Viswanath Poosala}, \bibinfo{person}{Peter~J
  Haas}, \bibinfo{person}{Yannis~E Ioannidis}, {and} \bibinfo{person}{Eugene~J
  Shekita}.} \bibinfo{year}{1996}\natexlab{}.
\newblock \showarticletitle{Improved histograms for selectivity estimation of
  range predicates}.
\newblock \bibinfo{journal}{\emph{ACM Sigmod Record}} \bibinfo{volume}{25},
  \bibinfo{number}{2} (\bibinfo{year}{1996}), \bibinfo{pages}{294--305}.
\newblock


\bibitem[\protect\citeauthoryear{Rezende, Mohamed, and Wierstra}{Rezende
  et~al\mbox{.}}{2014}]%
        {rezende2014stochastic}
\bibfield{author}{\bibinfo{person}{Danilo~Jimenez Rezende},
  \bibinfo{person}{Shakir Mohamed}, {and} \bibinfo{person}{Daan Wierstra}.}
  \bibinfo{year}{2014}\natexlab{}.
\newblock \showarticletitle{Stochastic backpropagation and approximate
  inference in deep generative models}. In \bibinfo{booktitle}{\emph{ICML}}.
\newblock


\bibitem[\protect\citeauthoryear{Riondato, Akdere, {\c{C}}etintemel, Zdonik,
  and Upfal}{Riondato et~al\mbox{.}}{2011}]%
        {riondato2011vc}
\bibfield{author}{\bibinfo{person}{Matteo Riondato}, \bibinfo{person}{Mert
  Akdere}, \bibinfo{person}{Uǧur {\c{C}}etintemel}, \bibinfo{person}{Stanley~B
  Zdonik}, {and} \bibinfo{person}{Eli Upfal}.} \bibinfo{year}{2011}\natexlab{}.
\newblock \showarticletitle{The VC-dimension of SQL queries and selectivity
  estimation through sampling}. In \bibinfo{booktitle}{\emph{ECML PKDD}}.
  Springer, \bibinfo{pages}{661--676}.
\newblock


\bibitem[\protect\citeauthoryear{Rudin et~al\mbox{.}}{Rudin
  et~al\mbox{.}}{1964}]%
        {rudin1964principles}
\bibfield{author}{\bibinfo{person}{Walter Rudin} {et~al\mbox{.}}}
  \bibinfo{year}{1964}\natexlab{}.
\newblock \bibinfo{booktitle}{\emph{Principles of mathematical analysis}}.
  Vol.~\bibinfo{volume}{3}.
\newblock \bibinfo{publisher}{McGraw-hill New York}.
\newblock


\bibitem[\protect\citeauthoryear{Rumelhart and McClelland}{Rumelhart and
  McClelland}{1987}]%
        {rumelhart1987learning}
\bibfield{author}{\bibinfo{person}{David~E Rumelhart} {and}
  \bibinfo{person}{James~L McClelland}.} \bibinfo{year}{1987}\natexlab{}.
\newblock \showarticletitle{Learning Internal Representations by Error
  Propagation}.
\newblock  (\bibinfo{year}{1987}).
\newblock


\bibitem[\protect\citeauthoryear{Salimans, Karpathy, Chen, and Kingma}{Salimans
  et~al\mbox{.}}{2017}]%
        {salimans2017pixelcnn++}
\bibfield{author}{\bibinfo{person}{Tim Salimans}, \bibinfo{person}{Andrej
  Karpathy}, \bibinfo{person}{Xi Chen}, {and} \bibinfo{person}{Diederik~P
  Kingma}.} \bibinfo{year}{2017}\natexlab{}.
\newblock \showarticletitle{Pixelcnn++: Improving the pixelcnn with discretized
  logistic mixture likelihood and other modifications}.
\newblock \bibinfo{journal}{\emph{ICLR}}.
\newblock


\bibitem[\protect\citeauthoryear{Schmidhuber}{Schmidhuber}{2015}]%
        {schmidhuber2015deep}
\bibfield{author}{\bibinfo{person}{J{\"u}rgen Schmidhuber}.}
  \bibinfo{year}{2015}\natexlab{}.
\newblock \showarticletitle{Deep learning in neural networks: An overview}.
\newblock \bibinfo{journal}{\emph{Neural networks}}  \bibinfo{volume}{61}
  (\bibinfo{year}{2015}), \bibinfo{pages}{85--117}.
\newblock


\bibitem[\protect\citeauthoryear{Scott}{Scott}{2015}]%
        {scott2015multivariate}
\bibfield{author}{\bibinfo{person}{David~W Scott}.}
  \bibinfo{year}{2015}\natexlab{}.
\newblock \bibinfo{booktitle}{\emph{Multivariate density estimation: theory,
  practice, and visualization}}.
\newblock \bibinfo{publisher}{John Wiley \& Sons}.
\newblock


\bibitem[\protect\citeauthoryear{Spiegel and Polyzotis}{Spiegel and
  Polyzotis}{2006}]%
        {spiegel2006graph}
\bibfield{author}{\bibinfo{person}{Joshua Spiegel} {and}
  \bibinfo{person}{Neoklis Polyzotis}.} \bibinfo{year}{2006}\natexlab{}.
\newblock \showarticletitle{Graph-based synopses for relational selectivity
  estimation}. In \bibinfo{booktitle}{\emph{SIGMOD}}.
  \bibinfo{pages}{205--216}.
\newblock


\bibitem[\protect\citeauthoryear{Stillger, Lohman, Markl, and Kandil}{Stillger
  et~al\mbox{.}}{2001}]%
        {stillger2001leo}
\bibfield{author}{\bibinfo{person}{Michael Stillger}, \bibinfo{person}{Guy~M
  Lohman}, \bibinfo{person}{Volker Markl}, {and} \bibinfo{person}{Mokhtar
  Kandil}.} \bibinfo{year}{2001}\natexlab{}.
\newblock \showarticletitle{LEO-DB2's learning optimizer}. In
  \bibinfo{booktitle}{\emph{VLDB}}, Vol.~\bibinfo{volume}{1}.
  \bibinfo{pages}{19--28}.
\newblock


\bibitem[\protect\citeauthoryear{Sun and Li}{Sun and Li}{2019}]%
        {sun2019end}
\bibfield{author}{\bibinfo{person}{Ji Sun} {and} \bibinfo{person}{Guoliang
  Li}.} \bibinfo{year}{2019}\natexlab{}.
\newblock \showarticletitle{An end-to-end learning-based cost estimator}.
\newblock \bibinfo{journal}{\emph{VLDB}} \bibinfo{volume}{13},
  \bibinfo{number}{3} (\bibinfo{year}{2019}), \bibinfo{pages}{307--319}.
\newblock


\bibitem[\protect\citeauthoryear{Thaper, Guha, Indyk, and Koudas}{Thaper
  et~al\mbox{.}}{2002}]%
        {thaper2002dynamic}
\bibfield{author}{\bibinfo{person}{Nitin Thaper}, \bibinfo{person}{Sudipto
  Guha}, \bibinfo{person}{Piotr Indyk}, {and} \bibinfo{person}{Nick Koudas}.}
  \bibinfo{year}{2002}\natexlab{}.
\newblock \showarticletitle{Dynamic multidimensional histograms}. In
  \bibinfo{booktitle}{\emph{SIGMOD}}. \bibinfo{pages}{428--439}.
\newblock


\bibitem[\protect\citeauthoryear{To, Chiang, and Shahabi}{To
  et~al\mbox{.}}{2013}]%
        {to2013entropy}
\bibfield{author}{\bibinfo{person}{Hien To}, \bibinfo{person}{Kuorong Chiang},
  {and} \bibinfo{person}{Cyrus Shahabi}.} \bibinfo{year}{2013}\natexlab{}.
\newblock \showarticletitle{Entropy-based histograms for selectivity
  estimation}. In \bibinfo{booktitle}{\emph{CIKM}}.
  \bibinfo{pages}{1939--1948}.
\newblock


\bibitem[\protect\citeauthoryear{Tzoumas, Deshpande, and Jensen}{Tzoumas
  et~al\mbox{.}}{2011}]%
        {tzoumas2011lightweight}
\bibfield{author}{\bibinfo{person}{Kostas Tzoumas}, \bibinfo{person}{Amol
  Deshpande}, {and} \bibinfo{person}{Christian~S Jensen}.}
  \bibinfo{year}{2011}\natexlab{}.
\newblock \showarticletitle{Lightweight graphical models for selectivity
  estimation without independence assumptions}.
\newblock \bibinfo{journal}{\emph{PVLDB}} \bibinfo{volume}{4},
  \bibinfo{number}{11} (\bibinfo{year}{2011}), \bibinfo{pages}{852--863}.
\newblock


\bibitem[\protect\citeauthoryear{Tzoumas, Deshpande, and Jensen}{Tzoumas
  et~al\mbox{.}}{2013}]%
        {tzoumas2013efficiently}
\bibfield{author}{\bibinfo{person}{Kostas Tzoumas}, \bibinfo{person}{Amol
  Deshpande}, {and} \bibinfo{person}{Christian~S Jensen}.}
  \bibinfo{year}{2013}\natexlab{}.
\newblock \showarticletitle{Efficiently adapting graphical models for
  selectivity estimation}.
\newblock \bibinfo{journal}{\emph{The VLDB Journal}} \bibinfo{volume}{22},
  \bibinfo{number}{1} (\bibinfo{year}{2013}), \bibinfo{pages}{3--27}.
\newblock


\bibitem[\protect\citeauthoryear{Van~Gelder}{Van~Gelder}{1993}]%
        {van1993multiple}
\bibfield{author}{\bibinfo{person}{Allen Van~Gelder}.}
  \bibinfo{year}{1993}\natexlab{}.
\newblock \showarticletitle{Multiple join size estimation by virtual domains}.
  In \bibinfo{booktitle}{\emph{Proceedings of the twelfth ACM
  SIGACT-SIGMOD-SIGART symposium on Principles of database systems}}.
  \bibinfo{pages}{180--189}.
\newblock


\bibitem[\protect\citeauthoryear{Wang, Shen, and Zhang}{Wang
  et~al\mbox{.}}{2005}]%
        {wang2005nonlinear}
\bibfield{author}{\bibinfo{person}{Qiang Wang}, \bibinfo{person}{Yi Shen},
  {and} \bibinfo{person}{Jian~Qiu Zhang}.} \bibinfo{year}{2005}\natexlab{}.
\newblock \showarticletitle{A nonlinear correlation measure for multivariable
  data set}.
\newblock \bibinfo{journal}{\emph{Physica D: Nonlinear Phenomena}}
  \bibinfo{volume}{200}, \bibinfo{number}{3-4} (\bibinfo{year}{2005}),
  \bibinfo{pages}{287--295}.
\newblock


\bibitem[\protect\citeauthoryear{Williams}{Williams}{1992}]%
        {williams1992simple}
\bibfield{author}{\bibinfo{person}{Ronald~J Williams}.}
  \bibinfo{year}{1992}\natexlab{}.
\newblock \showarticletitle{Simple statistical gradient-following algorithms
  for connectionist reinforcement learning}.
\newblock \bibinfo{journal}{\emph{Machine learning}} \bibinfo{volume}{8},
  \bibinfo{number}{3-4} (\bibinfo{year}{1992}), \bibinfo{pages}{229--256}.
\newblock


\bibitem[\protect\citeauthoryear{Wu, Jindal, Amizadeh, Patel, Le, Qiao, and
  Rao}{Wu et~al\mbox{.}}{2018}]%
        {wu2018towards}
\bibfield{author}{\bibinfo{person}{Chenggang Wu}, \bibinfo{person}{Alekh
  Jindal}, \bibinfo{person}{Saeed Amizadeh}, \bibinfo{person}{Hiren Patel},
  \bibinfo{person}{Wangchao Le}, \bibinfo{person}{Shi Qiao}, {and}
  \bibinfo{person}{Sriram Rao}.} \bibinfo{year}{2018}\natexlab{}.
\newblock \showarticletitle{Towards a learning optimizer for shared clouds}.
\newblock \bibinfo{journal}{\emph{VLDB}} \bibinfo{volume}{12},
  \bibinfo{number}{3} (\bibinfo{year}{2018}), \bibinfo{pages}{210--222}.
\newblock


\bibitem[\protect\citeauthoryear{Yang, Kamsetty, Luan, Liang, Duan, Chen, and
  Stoica}{Yang et~al\mbox{.}}{2021}]%
        {yang2020neurocard}
\bibfield{author}{\bibinfo{person}{Zongheng Yang}, \bibinfo{person}{Amog
  Kamsetty}, \bibinfo{person}{Sifei Luan}, \bibinfo{person}{Eric Liang},
  \bibinfo{person}{Yan Duan}, \bibinfo{person}{Xi Chen}, {and}
  \bibinfo{person}{Ion Stoica}.} \bibinfo{year}{2021}\natexlab{}.
\newblock \showarticletitle{NeuroCard: One Cardinality Estimator for All
  Tables}.
\newblock \bibinfo{journal}{\emph{PVLDB}} (\bibinfo{year}{2021}).
\newblock


\bibitem[\protect\citeauthoryear{Yang, Liang, Kamsetty, Wu, Duan, Chen, Abbeel,
  Hellerstein, Krishnan, and Stoica}{Yang et~al\mbox{.}}{2020}]%
        {naru}
\bibfield{author}{\bibinfo{person}{Zongheng Yang}, \bibinfo{person}{Eric
  Liang}, \bibinfo{person}{Amog Kamsetty}, \bibinfo{person}{Chenggang Wu},
  \bibinfo{person}{Yan Duan}, \bibinfo{person}{Xi Chen},
  \bibinfo{person}{Pieter Abbeel}, \bibinfo{person}{Joseph~M Hellerstein},
  \bibinfo{person}{Sanjay Krishnan}, {and} \bibinfo{person}{Ion Stoica}.}
  \bibinfo{year}{2020}\natexlab{}.
\newblock \showarticletitle{Deep Unsupervised Cardinality Estimation}.
\newblock \bibinfo{journal}{\emph{VLDB}} \bibinfo{volume}{13},
  \bibinfo{number}{3}, \bibinfo{pages}{279--292}.
\newblock


\bibitem[\protect\citeauthoryear{Zanettin}{Zanettin}{2019}]%
        {dmv}
\bibfield{author}{\bibinfo{person}{Federico Zanettin}.}
  \bibinfo{year}{2019}\natexlab{}.
\newblock \bibinfo{title}{State of New York. Vehicle, snowmobile, and boat
  registrations}.
\newblock
\newblock
\newblock
\shownote{\url{catalog.data.gov/dataset/vehicle-snowmobile-and-boat-registrations}.}


\bibitem[\protect\citeauthoryear{Zhao, Christensen, Li, Hu, and Yi}{Zhao
  et~al\mbox{.}}{2018}]%
        {zhao2018random}
\bibfield{author}{\bibinfo{person}{Zhuoyue Zhao}, \bibinfo{person}{Robert
  Christensen}, \bibinfo{person}{Feifei Li}, \bibinfo{person}{Xiao Hu}, {and}
  \bibinfo{person}{Ke Yi}.} \bibinfo{year}{2018}\natexlab{}.
\newblock \showarticletitle{Random sampling over joins revisited}. In
  \bibinfo{booktitle}{\emph{SIGMOD}}. \bibinfo{pages}{1525--1539}.
\newblock


\end{thebibliography}
\end{document}